\documentclass[review,12pt,authoryear]{elsarticle}
\usepackage{framed,multirow}
\usepackage{fullpage}
\usepackage{nomencl} 
\makenomenclature
\renewcommand*\nompreamble{\begin{multicols}{2}}
\renewcommand*\nompostamble{\end{multicols}}
\usepackage{amssymb}
\usepackage{amsmath}
\usepackage{graphicx}
\usepackage[retainorgcmds]{IEEEtrantools}
\usepackage[colorlinks]{hyperref}
\hypersetup{
  citecolor = {blue}
}
\usepackage{siunitx}
\usepackage{pdflscape}

\usepackage{enumitem}
\usepackage[normalem]{ulem}
\useunder{\uline}{\ul}{}
\usepackage{setspace} 
\usepackage[utf8]{inputenc}
\usepackage[linesnumbered, ruled, vlined]{algorithm2e}
\usepackage{rotating}
\usepackage{nicefrac}
\usepackage{xspace}
\usepackage{float}
\usepackage{caption}
\usepackage{subcaption}
\usepackage{lscape, longtable, tabu, booktabs, multirow}
\usepackage[capitalise, noabbrev]{cleveref}
\usepackage{verbatim}
\usepackage{mathtools}
\usepackage{makecell}
\usepackage{lineno}
\usepackage{threeparttable}
\makeatletter
\def\ps@pprintTitle{%
 \let\@oddhead\@empty
 \let\@evenhead\@empty
 \def\@oddfoot{}%
 \let\@evenfoot\@oddfoot}
\makeatother
\def\tsc#1{\csdef{#1}{\textsc{\lowercase{#1}}\xspace}}
\tsc{WGM}
\tsc{QE}
\tsc{EP}
\tsc{PMS}
\tsc{BEC}
\tsc{DE}

\begin{document}
\let\WriteBookmarks\relax
\def\floatpagepagefraction{1}
\def\textpagefraction{.001}

\title {Enhancing U.S. swine farm preparedness for infectious foreign animal diseases with rapid access to biosecurity information}

\author[1,2]{Christian Fleming}

\author[1]{Kelsey Mills}

\author[1]{Nicolas C. Cardenas}

\author[1]{Jason A. Galvis}

\author[3]{Cesar Corzo}

\author[1,2]{Gustavo Machado\corref{cor2}}
\ead{gmachad@ncsu.edu}

\cortext[cor2]{Corresponding Author.}

\affiliation[1]{organization={Department of Population Health and Pathobiology, College of Veterinary Medicine, North Carolina State University},
    addressline={1060 William Moore Drive}, 
    city={Raleigh},
    postcode={27607},
    state={NC},
    country={USA}}
\affiliation[2]{organization={Center for Geospatial Analytics, North Carolina State University},
    addressline={5112 Jordan Hall}, 
    city={Raleigh},
    postcode={27607}, 
    state={NC},
    country={USA}}
\affiliation[3]{organization={Department of Veterinary Population Medicine, College of Veterinary Medicine, University of Minnesota},
    addressline={1365 Gortner Avenue}, 
    city={St. Paul},
    postcode={55108}, 
    state={MN},
    country={USA}}

\begin{abstract} 
The U.S. Secure Pork Supply (SPS) Plan is a voluntary program that establishes biosecurity standards to maintain business continuity. The role of biosecurity in disease spread is well recognized, yet the U.S. swine industry lacks knowledge of individual farm biosecurity and the efficacy of existing measures.

Here, we (i) described a consortium among the swine industry, government, and academia that formed the Rapid Access Biosecurity (RAB) app\texttrademark\, (ii) summarized the farm characteristics and biosecurity of premises in RABapp\texttrademark\, (iii) mapped RABapp\texttrademark's biosecurity coverage of the U.S. swine population, and (iv) evaluated associations between biosecurity measures and reports of porcine reproductive and respiratory syndrome virus (PRRSV) and porcine epidemic diarrhea virus (PEDV) using mixed-effects logistic regression.

RABapp\texttrademark, used in 31 states, represented 42\% of the U.S. commercial swine population. 
76\% (234/307) of Agricultural Statistics Districts were identified as biosecurity deserts, with less than half of their swine population represented in RABapp\texttrademark. Requiring footwear or clothing changes, multiple carcass disposal locations, greater distance from neighboring swine premises, using bait for rodent control, and carcass burial significantly reduced the odds of PRRSV/PEDV occurrence. Conversely, rendering carcasses, deep pit or tank manure storage, land application of manure, and use of grounds-keeping for rodent control were associated with increased odds of infection.

This study demonstrated the national importance of RABapp\texttrademark\ as a centralized repository, mapped SPS plan adoption, curtailing high-risk practices, and reinforcing the identified measures could help reduce the circulation of endemic disease and strengthen industry preparedness for future foreign animal disease emergencies.

\end{abstract}


\maketitle

\section{Introduction} \label{sec:intro} 
\noindent On-farm biosecurity is the first and last defense against the introduction of pathogens in livestock populations \citep{jurado_risk_2019,silva_machine-learning_2019,alarcon_biosecurity_2021}.
When implemented effectively, biosecurity practices restrict the introduction of pathogens and reduce their spread, e.g., through personnel and fomites \citep{dixon_african_2020, galvis_betweenfarm_2022, jurado_risk_2019}.
Enhanced biosecurity plans that can be implemented promptly are essential in controlling the spread of infectious diseases, especially when a vaccine is not an option or when pharmaceutical interventions are not reliable; this is particularly important to prevent the introduction and dissemination of large-scale transboundary diseases such as African swine fever (ASF) \citep{mighell_african_2021,sykes_estimating_2023}.
Although animal and vehicle movements are the main drivers of disease spread between farms \citep{andraud_threat_2019,machado_modelling_2022,sykes_estimating_2023,galvis_role_2024}, restrictions on these movements work best to control outbreaks when they are supported by effective transmission-limiting practices, such as clearly defined line of separation segregating clean from contaminated areas  \citep{silva_machine-learning_2019,chenais_epidemiological_2019,mulumbamfumu_african_2019,olsevskis_african_2016,european_food_safety_authority_efsa_epidemiological_2024}. The efficacy of biosecurity plans and the types of biosecurity employed vary significantly between farms and types of farms \citep{alarcon_biosecurity_2021,silva_machine-learning_2019,campler_description_2024}. Producers can choose from various biosecurity measures to protect against pathogens that are often introduced into swine production due to deficits in biosecurity and human-mediated pathways, such as via fomites and shared equipment \citep{silva_machine-learning_2019,pudenz_adoption_2019,lee_swine_2021,lamberga_african_2020,deka_modeling_2024}.
On-farm biosecurity protects livestock not only from naturally occurring diseases but also from intentional threats, such as agroterrorism \citep{green_confronting_2019}.
Consequently, successful responses to future outbreaks depend on farms having effective biosecurity measures \citep{klein_biosecurity_2024, campler_description_2024}. Efforts towards establishing regional and national farm-level biosecurity programs are remarkably different during peacetime compared to times of adverse disease events \citep{wang_african_2018,li_african_2020}, the latter being by far more expensive and less effective \citep{liu_prevention_2021,klein_biosecurity_2024}.
The spread of ASF across Asia and Europe, and recently near the United States in the Dominican Republic \citep{gonzales_african_2021,schambow_update_2025} and Haiti \citep{jean-pierre_analysis_2022}, has shown how regional differences in on-farm biosecurity and production systems shape disease vulnerability, transmission, and impacts \citep{gonzales_african_2021,schambow_update_2025,jean-pierre_analysis_2022,mutua_context_2021,viltrop_biosecurity_2021,klein_biosecurity_2024}.

The U.S. swine industry would benefit from a collective understanding of individual-farm biosecurity plans and the efficacy of current measures in preventing the introduction of new pathogens. Such knowledge could also reduce the spread of endemic diseases such as porcine reproductive and respiratory syndrome virus (PRRSV) and porcine epidemic diarrhea virus (PEDV) \citep{pudenz_adoption_2019,campler_description_2024}. Thus, there is an urgent need to catalog and review the biosecurity measures of individual farms and prioritize effective measures before foreign animal diseases (FADs) are introduced into the U.S. \citep{campler_description_2024}. In addition, the
producers should have access to information about the benefits of implementing a biosecurity plan, as well as clear guidance and a thorough understanding of regulatory requirements. They should have information, at a minimum, on how each specific measure reduces the risk of disease introduction and spread, which would contribute to implementation decisions \citep{klein_biosecurity_2024,agrawal_evaluating_2024}.

In the U.S., Secure Food Supply (SFS) Plans have been developed for all food-animal systems, constituting a first-of-its-kind, voluntary, national biosecurity program that aims to keep healthy livestock and products moving, protect the food supply, and preserve business continuity across the production chain \citep{usda_aphis_veterinary_services_foreign_2016}.
In contrast, the European Animal Health Law assigns animal owners the responsibility for implementing biosecurity, with mandatory physical and management measures \citep{klein_biosecurity_2024}.
The suite of SFS Plans includes the pork-focused Secure Pork Supply (SPS) Plan, which was developed through a public–private collaboration among the National Pork Board (Pork Checkoff), USDA’s Animal and Plant Health Inspection Service (APHIS), and partner universities \citep{usda_aphis_veterinary_services_foreign_2016}.
The SPS Plan provides guidelines to help producers establish enhanced biosecurity plans, also known as SPS biosecurity plans \citep{pudenz_adoption_2019, machado_rapid_2023,campler_description_2024}.
In the U.S., commercial swine farms are not mandated to have an SPS biosecurity plan; however, it is required for US SHIP certification \citep{us_swine_health_improvement_plan_us_2025}. Additionally, in some states, it is necessary for movement permits to be approved by State Animal Health Officials (SAHOs) before being granted. Notably, producers who have an SPS biosecurity plan that has been reviewed and approved by SAHOs are likely to return to normal operations after an FAD outbreak more quickly \citep{pudenz_adoption_2019,mitchell_effects_2019}.
Additionally, standardized records of animal movements, when combined with farm geolocations in a centralized and harmonized repository, will expedite traceability and protect business continuity \citep{machado_rapid_2023,sykes_estimating_2023,cardenas_analyzing_2024,campler_description_2024,galvis_role_2024}.
A centralized biosecurity and traceability database is essential for the development of large-scale disease transmission models, which may uncover how on-farm biosecurity acts as a barrier to between-farm disease dissemination \citep{sykes_interpretable_2022}.
Such models have already uncovered disease transmission pathways \citep{sykes_estimating_2023,galvis_betweenfarm_2022,galvis_modelling_2022,deka_modeling_2024,sykes_identifying_2025}, and shed light on the role of animal and vehicle movement in swine disease dissemination \citep{galvis_role_2024, galvis_modeling_2022}.

Thus, the development of RABapp\texttrademark\ as a national biosecurity and traceability repository in the U.S. \citep{machado_rapid_2023} enables the testing and identification of the most effective on-farm biosecurity measures against endemic swine diseases at the national level. The availability of records on the occurrence of endemic diseases, such as PRRSV and PEDV \citep{perez_individual_2019}, when coupled with the centralized RABapp\texttrademark\ repository, enables the investigation of the association between biosecurity measures and disease introductions \citep{havas_assessment_2023,dee_improvements_2023}.
Therefore, in this study, we describe the unique multi-sector initiative that led to the creation of the RABapp\texttrademark\ consortium, establishing a platform that standardizes SPS biosecurity plan creation and audits, stores and enables analysis of animal and vehicle movement data, facilitates the development of disease transmission models and disease emergency response tools, and supports national programs such as the US Swine Health Improvement Plan (US SHIP) \citep{harlow_biosecurity_2024, us_swine_health_improvement_plan_us_2025}. In addition, we analyzed 5,514 SPS biosecurity plans from 31 U.S. states, describing and comparing farms and SPS biosecurity plans across geographic regions and production types; and examining the relationship between the biosecurity measures and the occurrence of endemic diseases, such as PRRSV and PEDV

\section{Materials and Methods} \label{sec:methods}

\subsection{The RABapp\texttrademark\ web-based application} \label{sec:RABapp} \noindent 
RABapp\texttrademark\ is a consortium of academic researchers, livestock producers, state animal health officials (SAHOs), and industry veterinarians.
The members of the RABapp\texttrademark\ consortium use the application to work together towards common goals: prepare the U.S. swine industry for large-scale FAD emergencies and understand how endemic diseases disseminate. The primary functions of RABapp\texttrademark\ are to serve livestock producers and SAHOs with a repository of on-farm biosecurity plans, establish traceability of animals, semen, and vehicles, centralize disease records, and integrate with disease transmission models \citep{sykes_identifying_2025, galvis_estimating_2025}.
These RABapp\texttrademark\ functionalities for producers and SAHOs are described in detail in the RABapp\texttrademark\ handbook \citep{machado_rapid_2023} and website (\href{https://machado-lab.github.io/RABapp-systems/}{RABapp\texttrademark\ systems}).

\begin{figure*}[!htb] 
    \centering \includegraphics[scale=.60]{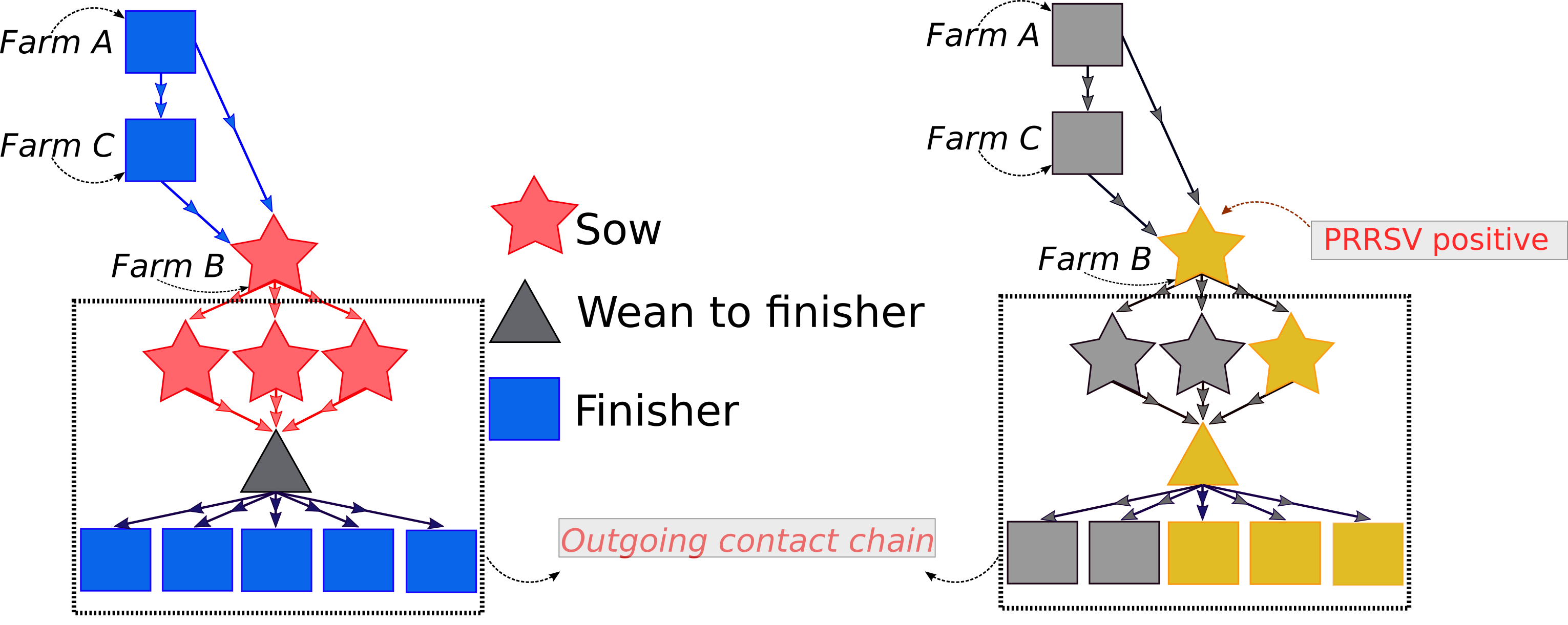}
    \caption{\emph{
        \textbf{Left:}
        This diagram shows an example of ingoing and outgoing contacts for a farm of interest (Farm B).
        Farm B has two farms within its ingoing contacts (Farm A and Farm C) and nine within its outgoing contacts (all farms within the gray square).
        \textbf{Right:}
        This diagram illustrates the same contacts, highlighting which farms have been infected with PRRSV (yellow) based on existing outbreak data.
    }} 
    \label{fig:contactrepeated}
    \end{figure*}

RABapp\texttrademark\ systems help swine, cattle, and poultry producers develop biosecurity plans. Once plans are completed according to national standards and guidelines (e.g., Secure Food Supply plans or The National Poultry Improvement Plan \citep{usda_aphis_veterinary_services_foreign_2016, us_poultry__egg_association_national_nodate}, they become available for SAHOs, who review, approve, or return them to the producers until the biosecurity plan satisfies their requirements.
The RABapp\texttrademark\ traceability functionality allows for the real-time tracing of direct and indirect farm-to-farm contact, allowing producers and SAHOs to track disease spread. 
Records of between-farm animal movements are received electronically and assessed for validity and completeness. At a minimum, they must include the source and destination premises, the movement date, and the number of units transported. Movement data with missing information is excluded and reported back to the participating producer on a regular basis.
In RABapp\texttrademark, the movement data creates a temporal contact network for traceability (Figure \ref{fig:contactrepeated}).
Identifying farms at high risk of outbreaks is possible, as the traceability functionality in RABapp\texttrademark\ displays the complete contact chain created by animal movements, including premises connected directly or indirectly via intermediary premises (Figure \ref{fig:contactrepeated}).

The RABapp\texttrademark\ control zone tool allows SAHOs to create and manage FAD incidents, as well as manage movement permits. Producers are notified when control zones are created and can track their pig, semen, and vehicle movements to and from these control zones. Thus, the tool identifies premises within the control area(s), collects and displays the SPS biosecurity plan status, and enables contact tracing by tracking movement in and out of premises.
In addition, RABapp\texttrademark\ tracks the infection status of each premises and overlays it on the animal-movement network, allowing producers and SAHOs to trace movements of infected pigs to or from farms of interest and take preventive action (Figure \ref{fig:contactrepeated}).

Finally, RABapp\texttrademark\ utilizes the premises, biosecurity, movement, and disease occurrence data to develop population transmission models, including farm-, barn-, and pen-level ASF transmission model \citep{sykes_estimating_2023,deka_modeling_2024,sykes_identifying_2025} and a PRRSV calibrated transmission model \citep{galvis_modelling_2022,galvis_modelling_2022-1}. These models are made available to the RABapp\texttrademark\ user via the Disease Surveillance Tool.

\subsection{RABapp\texttrademark\ premises} \label{sec:premises} \noindent
Farms in RABapp\texttrademark\ are defined by their nationally unique premises identification number (PIN).
The PIN identifies premises in the repository and links them to their corresponding biosecurity plan, movement data, outbreaks, and log of changes (e.g., biosecurity plan status and audit history).
For this study, premises data were retrieved from the RABapp\texttrademark\ repository containing data from 2019 to 2025.
Data included premises PIN, managing company, capacity (maximum number of animals housed on the premises), production type, geographic coordinates (latitude and longitude), address, descriptions of on-site businesses other than swine production, and biosecurity plan status
    \footnote{In RABapp\texttrademark\ SPS plan status can be
         \emph {Absent:} premises have been registered but have not yet initiated the creation of the biosecurity plan.
        \emph {Pending:} premises have initiated the creation of the biosecurity plan.
        \emph {Pre-Approved:} a SPS plan is completed in RABapp\texttrademark\, written and map is done.
        \emph {Needs Company Review:} Plans that have been reviewed by SAHOs but sent back to the producer with questions on the written plan section or maps, meaning that the plan was not approved and needs producer actions (making changes to the plans and re-routing the plan back to SAHO).
        \emph {Needs State Review:} Plans sent back to producers have been returned to SAHO for approval or rejection based on the producer's response to SAHO requests.
        \emph {Approved:} SAHO has completed their review and approved the plan.
        \citep{machado_rapid_2023}}.
Due to the wide range and regional variations in production types, we employed a string-matching approach to classify premises production types into one of nine categories: Sow, Boar Stud, Gilt, Farrow-to-Finish, Finisher, Nursery, Wean-to-Finish, Isolation, and Other.
These nine production type categories were further categorized into breeding (Sow, Boar Stud, Gilt, Farrow-to-finish), growing pig (Finisher, Nursery, Wean-to-finish), or other (Isolation, Other).
Geographic coordinates were used to calculate geodesic distances between sites, allowing us to determine both the distance to the nearest neighboring premises and the number of premises located within a 10 km radius.
The RABapp\texttrademark\ repository includes descriptions of businesses other than swine production that are present on each premises. Using a string-matching approach, these free-text descriptions were classified into one or more of five categories based on the presence of certain terms.
    1) Crops: captured descriptions containing 'crop', 'grain', 'hay', 'field', 'berry', 'home farm', and 'farming' while excluding 'grain storage', 'hay storage', and 'farming equipment'.
    2) Cattle: captured descriptions containing 'cattle', 'beef', 'cow', 'dairy', 'heifer', 'bovine', or 'steer'.
    3) Poultry: captured descriptions containing 'chicken', 'turkey', 'poultry', or 'duck'.
    4) Miscellaneous: encompassed 'storage', 'sheep', 'goat', 'horse', 'compost', 'feed', 'equipment', 'repair', and various other terms indicating operations not directly related to cultivation, cattle, or poultry.
    5) None: no description or an invalid description (e.g., the description described another swine production operation) was provided.
Finally, this data was used to calculate the distribution of production types, capacity, and the presence of businesses other than swine production for 7,781 premises.

\subsubsection{Biosecurity deserts} \label{sec:desert} \noindent
We defined “biosecurity deserts” as areas where a majority of the commercial swine population was not represented in the RABapp\texttrademark\ repository, then identified these areas by cross-referencing RABapp\texttrademark\ premises capacities with 2022 Census of Agriculture hog and pig inventory data \citep{national_agricultural_statistics_service_department_of_agriculture_2022_2024}.
For each ASD, we summed the pig capacities of all RABapp\texttrademark\ premises and compared that total with the aggregated county-level hog inventory.
ASDs were classified as a biosecurity desert when RABapp\texttrademark-represented capacity accounted for less than half of the district’s hog inventory. We mapped the ASDs to visualize the distribution of biosecurity deserts across the contiguous U.S. and the geographic coverage of the U.S. commercial swine population within RABapp\texttrademark.

\subsection{Secure Pork Supply biosecurity plans} \label{sec:sps_plans} 
\noindent SPS biosecurity plans comprise two types of information: i) written biosecurity plans created by veterinarians and/or farm managers and/or biosecurity coordinators, and ii) premises maps.
An SPS biosecurity plan is considered complete and receives Pre-Approved status in RABapp\texttrademark\ when the following information is present:
    1) premises PIN, company, location (address and coordinates), production type, and capacity (section \ref{sec:premises});
    2) written section with a complete set of responses (subsection \ref{written});
    3) aerial image of the premises marked with the required biosecurity map features (subsection \ref{map}).

\begin{figure*}[h]
    \centering \includegraphics[scale=.50]{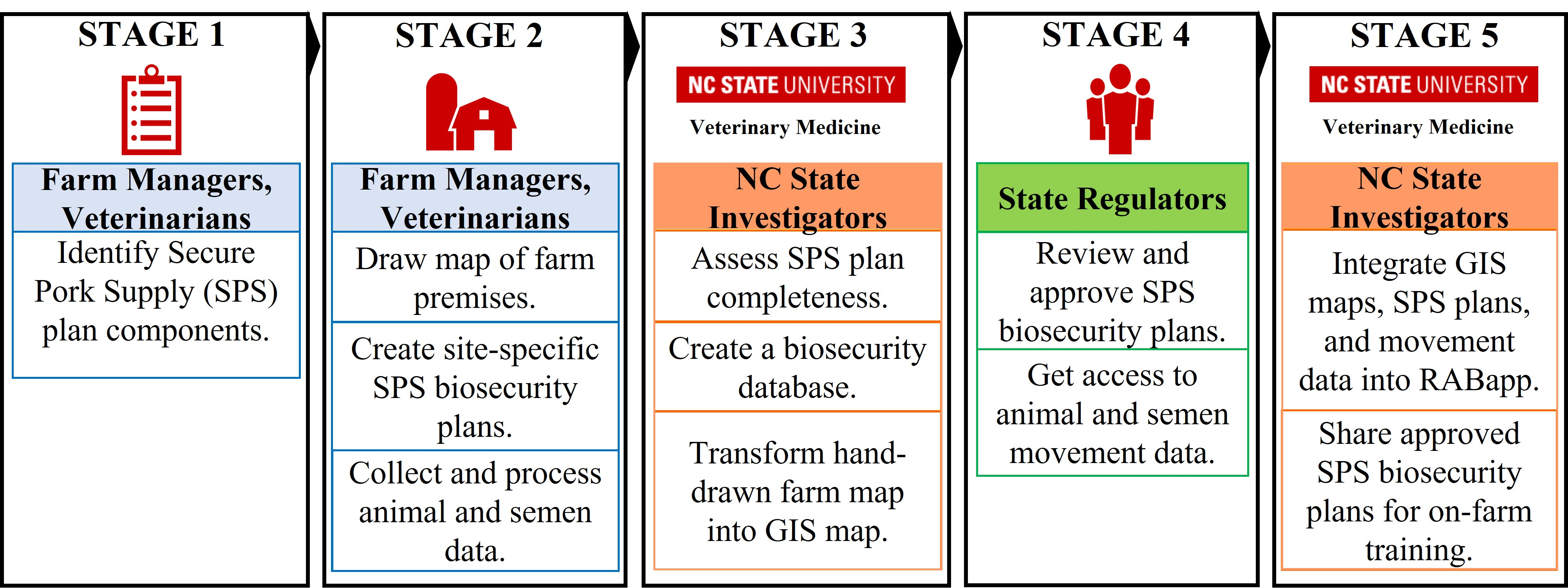}
    \caption{Conceptual stages of integrating biosecurity plans and animal movement data into RABapp\texttrademark}
    \label{fig:flow}
    \end{figure*}

Pork producers can add their SPS plans to RABapp\texttrademark\ and modify them as needed, while they are also available to SAHOs.
A series of stages is followed to ensure the strict and standardized cataloging, processing, and validation of SPS plans:
    Stage 1) The RABapp\texttrademark\ team fosters awareness and understanding of SPS biosecurity plan components among stakeholders.
    Stage 2) The RABapp\texttrademark\ team guides users in creating their SPS biosecurity plans, and as needed, generates SPS plan maps that are imported into RABapp\texttrademark. Otherwise, producers utilize the RABapp\texttrademark\ map maker tool to enter their own SPS plan maps directly into RABapp\texttrademark\ (https://machado-lab.github.io/RABapp-systems/rabappswine/map/).
    Stage 3) RABapp\texttrademark\ automatically assesses the SPS biosecurity plan for completeness as the producer enters it into RABapp\texttrademark.
    Stage 4) SAHOs review SPS biosecurity plans for approval and are granted access to movement data.
    Stage 5) The SPS biosecurity plans and combined movement data are readily available for SAHOs such that they can revise individual plans (Figure \ref{fig:flow}).
    
\subsubsection{Secure Pork Supply biosecurity plan written section} \label{written} \noindent
The written section captures information about on-farm biosecurity measures, infrastructure, and the contingency procedures for enhanced biosecurity during a FAD disease event \citep{secure_pork_supply_enahnced_nodate}. Information is provided by veterinarians, farm managers, or producers. The written section is divided into ten subsections: 1) Biosecurity manager, 2) Training, 3) Protecting the pig herd, 4) Vehicles and equipment, 5) Personnel, 6) Animal and semen movement, 7) Carcass disposal, 8) Manure management, 9) Pest control, and 10) Feed. We processed written section responses by systematically classifying free-text entries, while the items that are multiple choice or "yes"/"no" were retrieved as-is from the RABapp\texttrademark\ repository. For example, manure storage practices were classified using string-matching to capture mentions of lagoons ('lagoon', 'laguon', 'lagaon', 'pond', or 'earthen'), pits ('pit'), tanks ('tank', 'concrete', or 'slurry'), and land application ('spray', 'field', or 'broadcast').

\subsubsection{Secure Pork Supply biosecurity plan map} \label{map} \noindent
A premises biosecurity map displays the locations of biosecurity features overlaid on aerial imagery of the farm site. RABapp\texttrademark\ participants may submit completed maps to be imported or create maps directly in RABapp\texttrademark\ via the proprietary \href{https://machado-lab.github.io/RABapp-systems/rabappswine/map/}{map maker tool}.
To be considered complete, the maps must meet the minimum requirements outlined in the SPS biosecurity plan standards \citep{secure_pork_supply_enahnced_nodate}, which include information on the location of biosecurity features, such as the line of separation (LOS) that delineates live animals from the outside environment.
These are the required map features:
    1) Perimeter Buffer Area (PBA), an outer control boundary clearly marked around animal buildings to limit the movement of the virus near pig housing.
    2) PBA Access Point(s) (PBAAP), entry to the PBA is restricted to these controlled points, clearly marked with a sign and protected with a suitable barrier; 
    3) Line of Separation (LOS), a control boundary to prevent movement of virus into areas where susceptible animals can be exposed, formed in many situations by the walls of the building housing the animals; 
    4) LOS Access Point(s) (LOSAP), controlled points, clearly marked by a sign, where animals, people, or items cross the LOS and where equipment, people, and items follow specific biosecurity measures; 
    5) Cleaning and Disinfection (C\&D) Station, an operational, clearly marked, and equipped C\&D station with the means to remove visible contamination and then disinfect vehicles, equipment, and items needing to enter the PBA at a PBA Access Point; 
    6) Designated Parking Area (DPA), a clearly marked, designated parking area outside of the PBA, away from animal areas, for vehicles that will not enter the PBA and have not been cleaned and disinfected; 
    7) Carcass Disposal (CD), where dead animals are disposed of; 
    8) Carcass Removal Pathway (CRP), the route of carcass movement for disposal; 
    9) Vehicle Movement (VM), the route of vehicle movement (animal transport vehicles, deliveries, etc.); 
    and
    10) Site Entry (SE), entry to the site is restricted to these points that are protected with a gate or suitable barrier restricting access of unauthorized vehicles to the pork production facilities, and with signage conveying the restricted access. 
In addition to the required features above, RABapp\texttrademark\ users have the option to include the following non-mandatory features:
    11) Loading Chute (LC), the path of arriving animals as they are being loaded into the LOS; 
    12) Supply Area (SA), the area where deliveries are made outside of the PBA; 
    13) General Cleaning and Disinfection (GCD) stations for the cleaning and disinfection of vehicles, personnel, or equipment; 
    and
    14) Perimeter Buffer Area Animal Entrance (PBAAE) points designating a PBAAP as an animal loading/unloading area not used for a people entry point. 
    15) LOS Animal Emergency (LOSAE): This is an optional item that indicates which doors will be used to remove dead animals.
    16) Trash Dumpster (TD): This is an optional item that symbolizes any trash collection box.
The digital premises biosecurity map of each premises was retrieved from the RABapp\texttrademark\ repository and used to calculate the area and perimeter of the LOS, determine if the DPA, CD, or VM are within the PBA, and count the number of SE, LOSAP, LOS, and CD.

\begin{figure*}[!htb]  
    \centering \includegraphics[scale=.40]{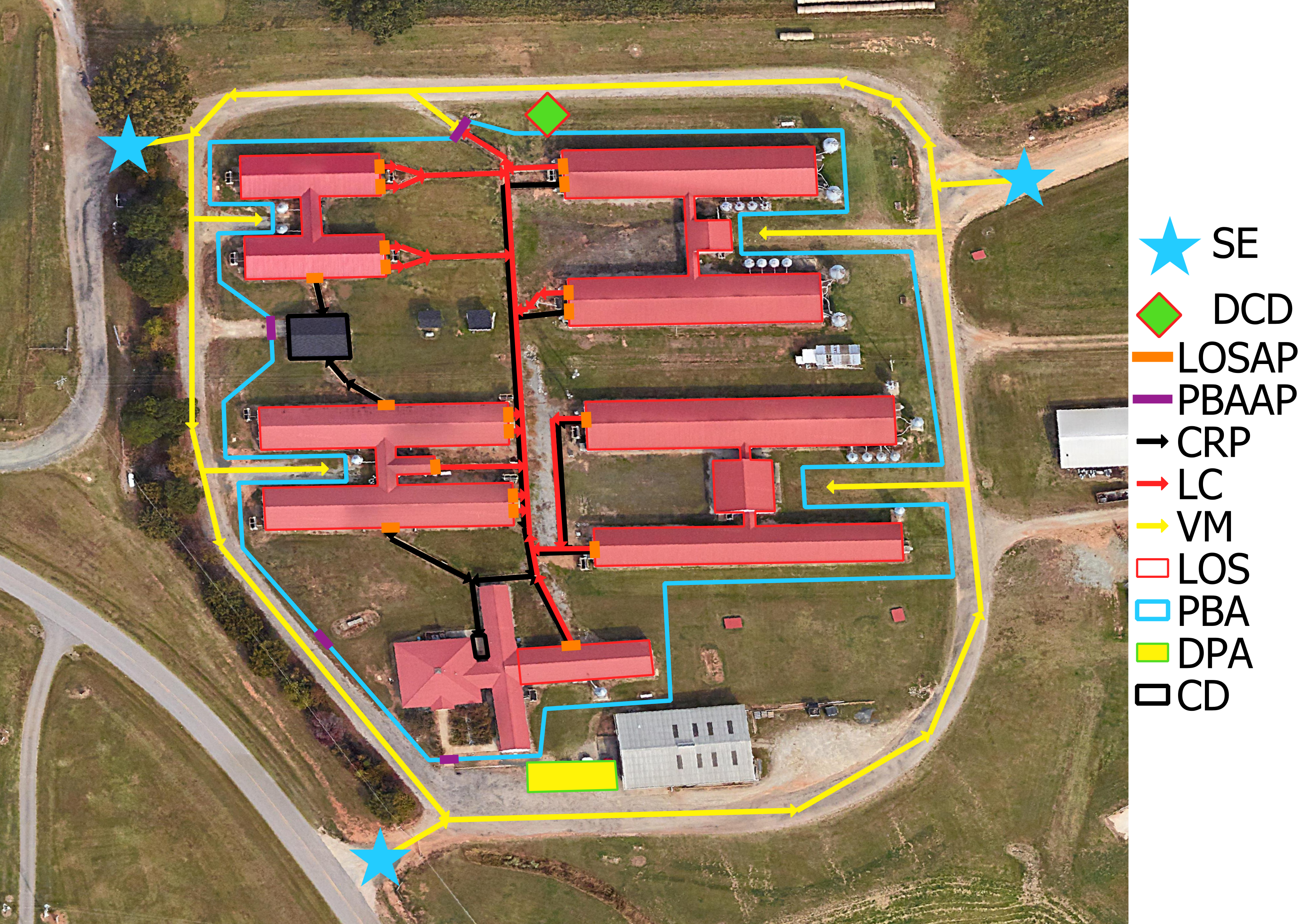}
    \caption{
        Secure Pork Supply biosecurity plan map with a legend and all features required by the SPS biosecurity plan, including the
        1) Site Entry (SE), symbolized by a blue star;
        2) Designated Cleaning and Disinfection Station (DCD), symbolized by a green diamond with red outline;
        3) Line of Separation Access Point (LOSAP), symbolized with an orange line on the LOS;
        4) Perimeter Buffer Area Access Point (PBAAP), symbolized with a purple line on the PBA;
        5) Carcass Disposal Pathway (CRP), symbolized with a black arrow;
        6) Loading Chute (LC), symbolized with a red arrow;
        7) Vehicle Movement (VM), symbolized with a yellow arrow;
        8) Line of Separation (LOS), symbolized with a red polygonal boundary;
        9) Perimeter Buffer Area (PBA), symbolized with a light blue boundary fully enclosing the LOS;
        10) Designated Parking Area (DPA), symbolized by a yellow rectangle with a green outline; and
        11) Carcass Disposal (CD), symbolized with a black rectangle.
    } 
    \label{fig:sps_map}
    \end{figure*}

\subsection{Disease occurrence data} \label{sec:outbreaks} \noindent
There are two ways in which a disease outbreak is shared with RABapp\texttrademark. Participants can share information on occurrences of PRRSV and PEDV, which includes the date of detection. The second route is from breeding farms participating in the Morrison Swine Health Monitoring Project (MSHMP), as RABapp\texttrademark\ is an MSHMP collaborator \citep{perez_individual_2019}.
In RABapp\texttrademark, premises infection status, as determined by these disease occurrence data, is color-coded throughout RABapp\texttrademark, shown in Figure \ref{fig:contactrepeated}.

\subsection{Biosecurity and disease outbreaks} \label{sec:model} \noindent
We modeled the probability that premises $i$ experienced at least one PRRSV or PEDV outbreak from January 1st, 2024, to December 31st, 2024 using a mixed-effects logistic regression:
\begin{equation}
\label{eq:model}
\begin{aligned}
  Y_i &\sim \text{Bernoulli}(p_i),\\[2pt]
  \text{logit}(p_i) &=
      \alpha_{j[i]} + \alpha_{k[i]} + \mathbf{x}_{i}^{\!\top}\boldsymbol{\beta},\\[2pt]
  \alpha_{j} &\sim \mathcal{N}\!\bigl(0,\sigma^{2}_{\text{company}}\bigr),
      & j&=1,\dots,J,\\
  \alpha_{k} &\sim \mathcal{N}\!\bigl(0,\sigma^{2}_{\text{breeding}}\bigr),
      & k&=1,\dots,K.
\end{aligned}
\end{equation}

Here, $Y_i$ is a binary outcome indicating whether premises $i$ reported an outbreak, $\mathbf{x}_{i}$ is the vector of biosecurity predictors, $\alpha_{j}$ is a random intercept for the managing company, and $\alpha_{k}$ is a random intercept for breeding versus growing-pig farms. This structure accounts for variability due to differences in operation and approach to biosecurity across companies, as well as between breeding and growing pig production.
We fitted this model using the $glmer$ function from the lme4 package in R \citep{bates_fitting_2015}.
We limited our modeling population to premises with a complete SPS plan (including the accompanying biosecurity plan map) from two companies that consistently shared disease occurrence data. 
There were 27 variables considered for model selection: 18 dichotomous variables, five categorical variables, and four continuous numeric variables (Table \ref{tab:results}).
Each of the 27 candidate variables were offered individually to the model, and only those with p \textless\ 0.20 were retained.
Multicollinearity was evaluated with three diagnostics: (i) Pearson $r$ for numeric predictor pairs, (ii) $\Phi$ (Pearson $\chi^2$) for binary-factor pairs, and (iii) variance-inflation factors (VIF).
Predictors with $|r|$ or $|\Phi|$ greater than $0.7$ or VIF greater than $5$ were removed one at a time, dropping the less statistically significant variable of each correlated pair; if the pair had equal univariate significance, the less biologically relevant was dropped.
With the set of 17 remaining candidate variables, we performed stepwise backward selection of predictors, dropping the least significant predictor (the one with the highest p-value) at each step and refitting the model until all predictors were significant at $\alpha=0.05$ (Supplementary Table S1).
For comparison, we also fitted a null model including only the random intercepts (company and breeding status) and no fixed effects.
For each predictor retained in the final model, we exponentiated the estimated regression coefficient to obtain the adjusted odds ratio (OR) and derived 95\% confidence intervals (CIs) and p-values from the model output, accounting for the random intercepts for company and breeding status.
We assessed the fit of the final model using the Akaike information criterion (AIC) and the area under the receiver operating characteristic (ROC) curve. The ROC curve was generated from the fitted outbreak probabilities of the final model using the $roc$ function from the pROC package \citep{robin_proc_2011}, providing a measure of the model's ability to discriminate between premises with and without outbreaks.
In addition, we conducted a chi-squared test for independence to compare outbreak frequencies between premises located in biosecurity deserts and those outside, limited to the subset of premises included in the regression model.

\subsection{Software} \label{sec:tools} \noindent
Data extraction and processing was performed in the Python (3.12.5) programming language using these libraries: sqlalchemy \citep{bayer_sqlalchemy_2012}, pandas \citep{the_pandas_development_team_pandas-devpandas_2024,mckinney_data_2010}, geopandas \citep{joris_van_den_bossche_geopandasgeopandas_2024}, geopy \citep{geopy_development_team_geopy_2023}.
Descriptive analysis and statistical modeling were conducted using R Statistical Software (v4.2.3) \citep{r_core_team_r_2024} using these packages: data.table \citep{barrett_datatable_2024}, lme4 \citep{bates_fitting_2015}, car \citep{fox_r_2019}, ggplot2 \citep{wickham_ggplot2_2016}, equatiomatic \citep{anderson_equatiomatic_2024}, pROC \citep{robin_proc_2011}.

\section{Results} \label{sec:results}

\subsection{Premises demographics and distribution} 
\begin{figure*}[h] \centering \includegraphics[scale=.50]{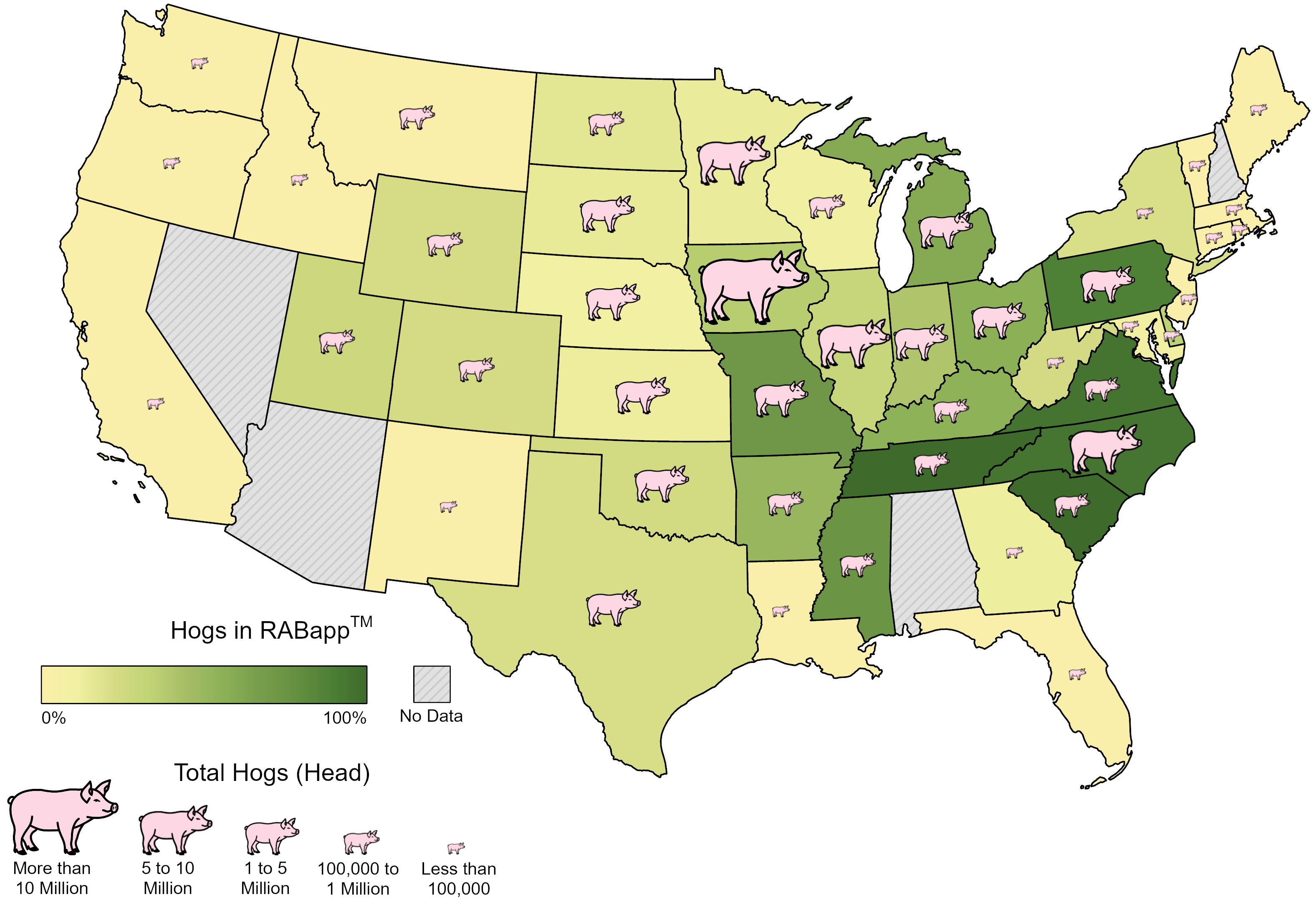}
    \caption{\emph{Total capacity, in head of swine, of the RABapp\texttrademark\ premises in each state as a proportion of the December 31, 2022 total hog inventory of that state as reported in the 2022 Census of Agriculture \citep{national_agricultural_statistics_service_department_of_agriculture_2022_2024}}}
    \label{fig:state}
    \end{figure*}

\noindent
At the time of writing this manuscript, RABapp\texttrademark, used by SAHOs and producers in 31 states, represented 42\% of the U.S. commercial swine population (Figure \ref{fig:state}). A total of 7,781 premises managed by 64 production companies were included in this study, with the highest numbers located in Iowa and North Carolina.
39\% of premises were of the wean-to-finish production type, followed by finisher at 38\%, nursery 11\%, and sow 9\%; other production types together accounted for less than 3\% of premises (Table \ref{tab:premises}).
The overall median capacity was 3,300 head (IQR: 2,400–4,955), with farrow-to-finish premises exhibiting the highest median capacity, at 5,500 head, among the nine production types. 28\% of premises reported having another business in addition to swine production---most frequently crops (25\%) or cattle (5\%). Of the nine production types, nursery premises most frequently had an additional business, with 45\% of premises reporting some business other than swine production. Wean-to-finish premises were least frequently associated with an additional business (11\%) (Table \ref{tab:premises}).

\begin{table}[!h]  \centering
    \caption{Distribution of production types, capacity, and the presence of a business other than swine production for 7,781 premises across 31 U.S. states.}
    \begin{threeparttable}
    \resizebox{\textwidth}{!}{%
    \begin{tabular}{@{}lllllllll@{}}
    \toprule
    Production type  & Count (\%)   & Capacity            &  & \multicolumn{5}{l}{Premises hosting business other than swine production (\%)}
                     \\ \cmidrule(lr){3-3} \cmidrule(l){5-9} 
                     &              & Median (IQR)        &  & Any business & Crops        & Cattle    & Poultry  & Misc.\textsuperscript{\dag}
                     \\ \midrule
    Wean-to-finish   & 3,058 (39\%) & 3,300 (2,460-4,960)   &  & 336 (11\%)   & 272 (9\%)    & 62 (2\%)  & 10 (\textless1\%) & 50 (2\%)  \\
    Finisher         & 2,954 (38\%) & 3,290 (2,400-4,859)   &  & 1,157 (39\%) & 1,043 (35\%) & 222 (8\%) & 47 (2\%)  & 34 (1\%)  \\
    Nursery          & 844 (11\%)   & 4,155 (2,600-6,400)   &  & 383 (45\%)   & 338 (40\%)   & 68 (8\%)  & 23 (3\%)  & 18 (2\%)  \\
    Sow              & 720 (9\%)    & 2,600 (1,800-4,014)   &  & 289 (40\%)   & 247 (34\%)   & 49 (7\%)  & 3 (\textless1\%)  & 34 (5\%)  \\
    Gilt             & 59 (1\%)     & 2,200 (1,200-4,020)   &  & 8 (14\%)     & 8 (14\%)     & 1 (2\%)   & 0         & 0         \\
    Farrow-to-finish & 57 (1\%)     & 5,500 (3,650-6,614)   &  & 8 (14\%)     & 6 (11\%)     & 2 (4\%)   & 0         & 2 (4\%)   \\
    Boar stud        & 37 (\textless1\%)    & 245 (178-360)         &  & 11 (30\%)    & 9 (24\%)     & 0         & 2 (5\%)   & 0         \\
    Other            & 33 (\textless1\%)    & 624 (160-1,000)       &  & 12 (36\%)    & 6 (18\%)     & 9 (27\%)  & 0         & 2 (6\%)   \\
    Isolation        & 19 (\textless1\%)    & 700 (375-1,000)       &  & 4 (21\%)     & 4 (21\%)     & 2 (11\%)  & 0         & 0         \\ \midrule
    Total            & 7,781        & 3,300 (2,400-4,950)   &  & 2,208 (28\%) & 1,933 (25\%) & 415 (5\%) & 85 (1\%)  & 140 (2\%) \\ \bottomrule
    \end{tabular} }
    \begin{tablenotes}[flushleft]
    \item[\dag] \small Miscellaneous: Any non-swine business other than cultivation, cattle, or poultry.
    \end{tablenotes}
    \end{threeparttable}
    \label{tab:premises}
    \end{table}

\subsubsection{Biosecurity deserts} 
\begin{figure*}[!htb]  \centering \includegraphics[scale=.60]{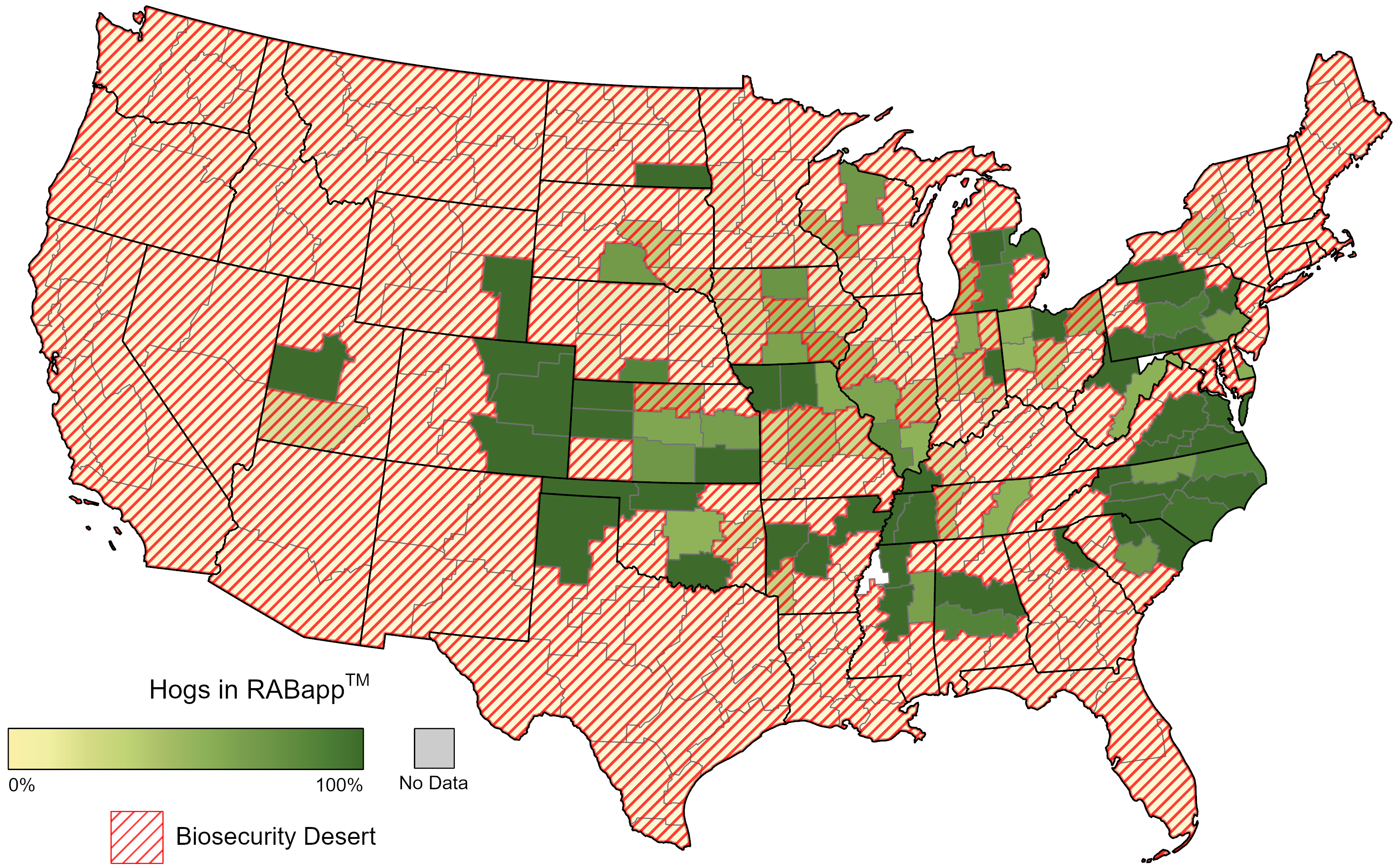}
    \caption{This map shows in green the total capacity of RABapp\texttrademark\ premises as a percentage of the total inventory of hogs as reported in the 2022 Census of Agriculture for each agricultural statistic district \citep{united_states_department_of_agriculture_2022_2024}. In 44 Agricultural Statistics Districts, the total capacity of RABapp\texttrademark\ premises exceeded the total inventory of hogs reported in the 2022 Census of Agriculture. Red hash marks are overlaid on districts meeting our criteria for a biosecurity desert (total capacity of RABapp\texttrademark\ premises is less than half of total commercial hog inventory).}
    \label{fig:deserts}
    \end{figure*}

\noindent
In Figure \ref{fig:deserts}, the red hatching depicts ASDs meeting the biosecurity desert criterion. Of 307 ASDs with commercial hog inventories, 234 (76\%) were identified as biosecurity deserts, encompassing 60.3\% of the national hog inventory. 31.5\% of RABapp\texttrademark\ premises were in a biosecurity desert; these premises had a total capacity of 11.47 million head. 

\subsection{SPS Biosecurity Plan} 
\begin{figure*}[!h] 
    \includegraphics[scale=.80]{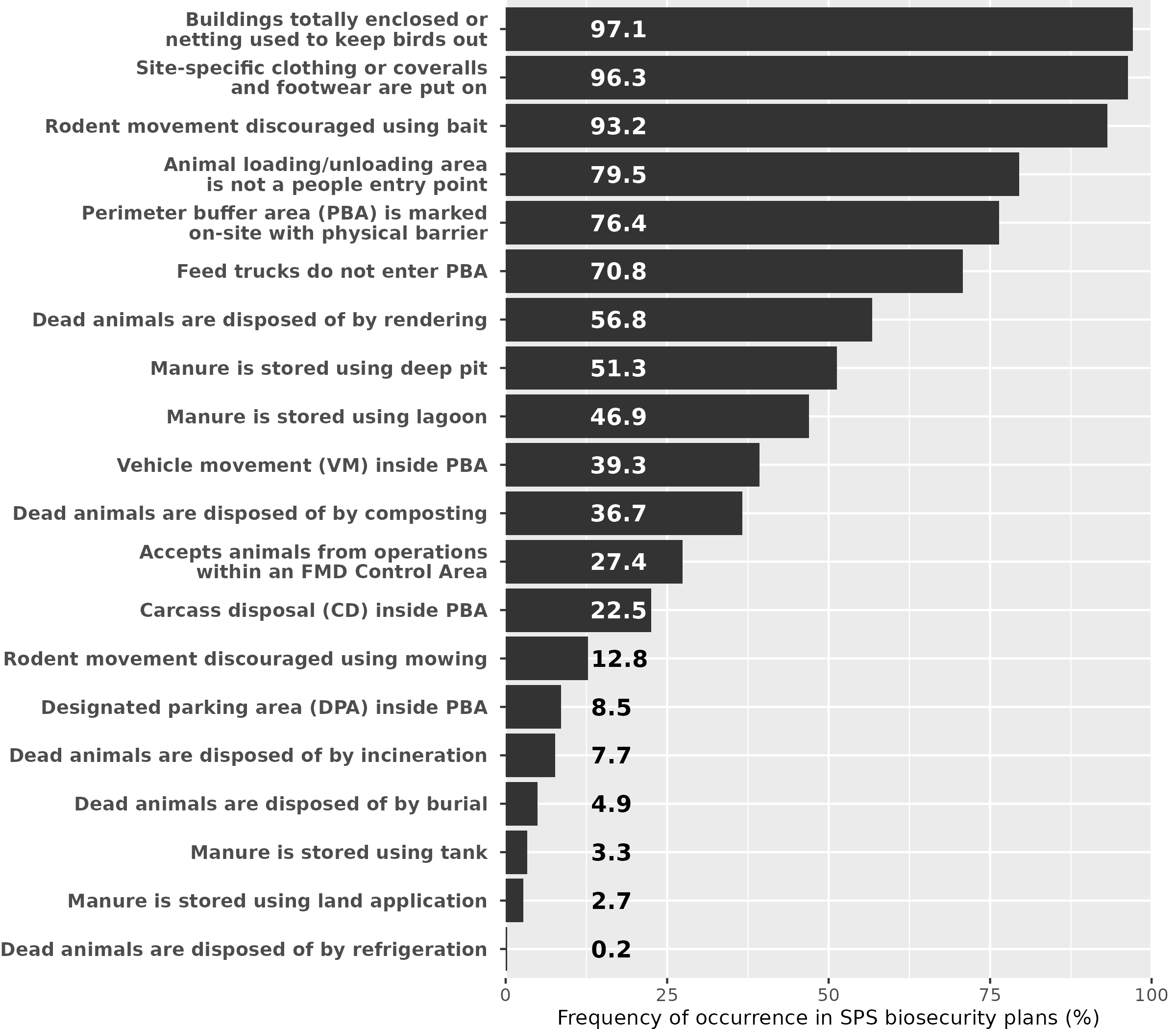}
    \caption{Distribution of biosecurity measures within 5,515 Secure Pork Supply biosecurity plans from the RABapp\texttrademark\ repository. The vertical axis lists individual biosecurity practices, and the horizontal axis indicates the percentage of biosecurity plans that include each practice.}
    \label{fig:spsbar}
    \end{figure*}

\noindent
Among the 7,781 premises, 5,515 (71\%) had a complete SPS biosecurity plan recorded in RABapp\texttrademark.
Some biosecurity practices were very widely implemented: 97\% of plans prevented birds from entering buildings with bird netting or buildings that are totally enclosed, and 96\% of plans required a footwear or clothing change as part of entry procedures (Figure \ref{fig:spsbar}).

Carcass disposal relied mainly on rendering (58\%) or composting (35\%). Incineration (8\%), burial (5\%), and refrigeration (\textless1\%) were infrequent. Manure storage was most commonly deep pit (50\%) or lagoon (48\%); tank storage (3\%) and land application (3\%) were rare.
The two most common methods of manure storage, lagoon and deep pit, emerged as alternative choices to the main method of manure storage. Specifically, 96\% of SPS plans included either a lagoon or a deep pit as a method of manure storage, while only 4\% (2\%) of SPS plans included both methods.
Nearly all sow farms had lagoons (89\%), and they were common on boar stud (83\%), farrow-to-finish (80\%), and nursery (69\%) premises, but were uncommon on wean-to-finish (21\%) premises.
Conversely, deep pit storage was prevalent on wean-to-finish premises (69\%) but infrequent on sow (17\%), boar stud (17\%), nursery (32\%), and farrow-to-finish (36\%) premises.

\begin{figure*}[!h]  \centering \includegraphics[scale=.8]{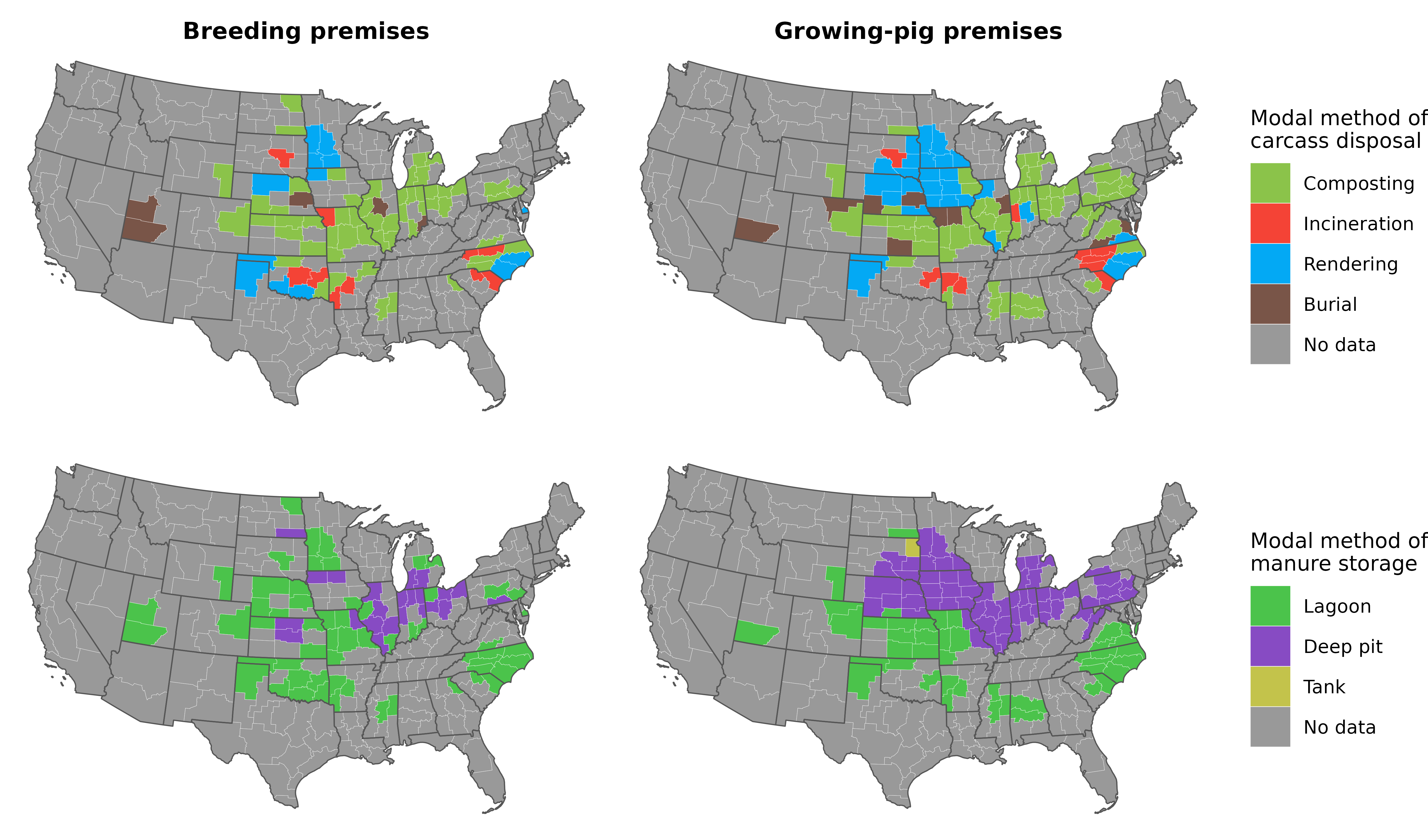}
    \caption{These maps show the distribution of manure management and carcass disposal methods in the SPS biosecurity plans of breeding and growing-pig premises within each Agricultural Statistics District.}
    \label{fig:4panel}
    \end{figure*}

Figure \ref{fig:4panel} shows the most frequent method of manure storage and carcass disposal for the breeding and growing-pig premises in each ASD.
Carcass disposal methods varied geographically, with rendering being more common in the western U.S. (from western Iowa and Minnesota westward) and southeastern North Carolina, while composting was predominant in eastern Iowa, Missouri, and regions farther east.
Despite rendering and composting being the most common carcass disposal methods overall (Figure \ref{fig:spsbar}), a considerable number of ASDs had burial or incineration as the most common method. However, ASDs within the highest swine density regions still tended to favor rendering.
The spatial distribution of the modal manure storage method for growing pig premises reveals a clear separation, with deep pit storage being more prevalent in the north and lagoons being more common in the south.
Among breeding premises, northern premises showed more preference for lagoon manure storage than growing pig premises.

We analyzed biosecurity maps that accompanied the 5,514 complete SPS plans.
Most biosecurity map features (SE, LOS, CD, cleaning stations) had a median count of one. The overall median LOS perimeter length was 467 m (IQR: 266-727), while the median LOS perimeter for sow farms was 1,174 m (IQR: 821–1,856).
Carcass disposal (CD) locations were within the PBA on 23\% of maps; occurring most frequently on boar stud (41\%) and farrow-to-finish (36\%) premises and least frequently on nursery (18\%), other (12\%), and isolation (0\%) premises.

\subsection{Biosecurity and disease outbreaks} \noindent
We used mixed-effects logistic regression with premises (n = 2,475) from two companies and with complete data; 889 of these premises reported $\geq 1$ outbreak of PRRSV or PEDV in 2024.
Three predictors were excluded for high collinearity ($\lvert r \rvert > 0.7$ or $VIF > 5$): 'buildings fully enclosed or bird netting used', 'manure is stored using lagoon', and 'dead animals are disposed of by composting'.
The univariable analysis identified 15 variables that were subsequently included in the multivariable model selection (Table \ref{tab:results}).
The final multivariable mixed-effects logistic regression model indicated that carcass rendering (OR: 3.52; CI: 2.69-4.6; p \textless\ 0.001), deep pit manure storage (OR: 2.04; CI: 1.41–2.95; p \textless\ 0.001), tank manure storage (OR: 1.73; CI: 1.07–2.81; p = 0.03), land application of manure (OR: 7.11; CI: 4.91–10.29; p \textless\ 0.001), and groundskeeping for rodent control (OR: 6.68; CI: 4.25–10.50; p \textless\ 0.001) were linked to significantly higher odds of PRRSV/PEDV outbreak.
Conversely, SPS plans requiring changes in footwear or clothing had significantly reduced odds of an outbreak (OR: 0.44; CI: 0.32–0.61; p \textless\ 0.001). The increase in distance to the nearest premises was also associated with reduced outbreak odds (OR per km: 0.92; CI: 0.88–0.95; p \textless\ 0.001), as was carcass burial (OR: 0.17; CI: 0.05–0.55; p = 0.003), and the use of bait for rodent control (OR: 0.21; CI: 0.15–0.31; p \textless\ 0.001).
The final model also included one significant predictor from the SPS biosecurity maps: the number of carcass disposal (CD) locations, which decreased the odds of infection (OR: 0.89; CI: 0.8–0.99; p = 0.036).
The final mixed-effects logistic regression model achieved an Akaike information criterion (AIC) of 2645.8, better than the null model (AIC = 3173.2) and the model before backward selection (AIC = 2648.6).
The area under the receiver operating characteristic (ROC) was 0.78.
In comparison, the null model achieved an area under the ROC (AUC) of 0.57, and the model before backward selection achieved an AUC of 0.79, demonstrating improved performance throughout the selection process.
Finally, the unadjusted chi-squared test comparing outbreak frequencies between premises located in biosecurity deserts and those outside indicated significantly higher outbreak frequencies among the premises in biosecurity deserts ($\chi^{2}$ = 35.9, p \textless\ 0.001).

\begin{table}[]
\centering
\caption{Estimated odds ratio and p-values of the fixed effects on biosecurity features associated with the occurrence of PRRSV or PEDV of 2,475 premises}
\begin{threeparttable}
\resizebox{\textwidth}{!}{%
\renewcommand{\arraystretch}{1.5}
\begin{tabular}{llllllllll}
\hline
 & \textit{Premises without outbreak ($N=1,586$)} & & \textit{Premises with outbreak ($N=889$)} & & \multicolumn{2}{l}{Univariable analysis} &  & \multicolumn{2}{l}{Multivariable analysis} \\ \cline{2-2} \cline{4-4} \cline{6-7} \cline{9-10} 
Variable & Median (IQR) & & Median (IQR) & & p-value & odds ratio (95\% CI) &  & odds ratio (95\% CI) & p-value \\ \hline
Distance to nearest premises (km) & 1.23 (0.43-2.69) & & 1.47 (0.76-2.49) & & \textless0.001 & 0.93 (0.90-0.96) &  & 0.92 (0.88-0.95) & \textless0.001 \\
Premises within 10 km & 17 (4-56) & & 19 (8-37) & & 0.53 & 1.001 (0.998-1.004) &  &  &  \\
Capacity & 3,800 (2,600-5,867) & & 4,111 (2,640-5,000) & & 0.22 & 0.98 (0.96-1.01)\textsuperscript{\dag} &  &  &  \\
Number of SE & 1 (1-1) & & 1 (1-1) & & 0.34 & 1.14 (0.87-1.495) &  &  &  \\
Number of LOSAP & 5 (3-8) & & 6 (4-9) & & 0.296 & 1.01 (0.99-1.03) &  &  &  \\
Number of LOS & 2 (1-3) & & 1 (1-2) & & \textless0.001 & 0.88 (0.84-0.93) &  &  &  \\
Number of CD & 1 (1-1) & & 1 (1-1) & & 0.048 & 0.91 (0.84-0.999) &  & 0.89 (0.8-0.99) & 0.036 \\
LOS area & 3,365 m\textsuperscript{2} (2,135-6,322) & & 3,382 m\textsuperscript{2} (2,136-4,753) & & 0.91 & 0.999 (0.98-1.02)\textsuperscript{\ddag} &  &  &  \\
LOS perimeter & 593 m (352-1091) & & 543 m (327-791) & & 0.47 & 1.01 (0.99-1.02)\textsuperscript{\S} &  &  &  \\ \hline
Variable & $n$ (\%) & & $n$ (\%) & & \multicolumn{2}{l}{Univariable analysis} & & \multicolumn{2}{l}{Multivariable analysis} \\ \hline
Has crops business & 977 (62\%) & & 330 (37\%) & & \textless0.001 & 0.42 (0.35-0.51) &  &  &  \\
Has cattle business & 143 (9\%) & & 63 (7\%) & & 0.58 & 0.92 (0.67-1.25) &  &  &  \\
Has poultry business & 36 (2\%) & & 7 (1\%) & & 0.03 & 0.40 (0.18-0.91) &  &  &  \\
Has misc. business & 26 (2\%) & & 13 (1\%) & & 0.94 & 0.97 (0.495-1.92) &  &  &  \\
Buildings totally enclosed or netting used to keep birds out & 1,527 (96\%) & & 789 (89\%) & & \textless0.001 & 0.25 (0.18-0.35) &  &  &  \\
Accepts animals from operations within a Control Area & 24 (2\%) & & 9 (1\%) & & 0.08 & 0.49 (0.22-1.08) &  &  &  \\
Site-specific clothing or coveralls and footwear are put on & 1,470 (93\%) & & 807 (91\%) & & 0.008 & 0.67 (0.495-0.90) &  & 0.44 (0.32-0.61) & \textless0.001 \\
Dead animals are disposed of by composting & 332 (21\%) & & 120 (13\%) & & \textless0.001 & 0.34 (0.26-0.44) &  &  &  \\
Dead animals are disposed of by incineration & 108 (7\%) & & 12 (1\%) & & \textless0.001 & 0.22 (0.12-0.398) &  &  &  \\
Dead animals are disposed of by rendering & 1,029 (65\%) & & 754 (85\%) & & \textless0.001 & 4.76 (3.72-6.07) &  & 3.52 (2.69-4.6) & \textless0.001 \\
Dead animals are disposed of by burial & 131 (8\%) & & 3 (\textless1\%) & & \textless0.001 & 0.04 (0.01-0.13) &  & 0.17 (0.05-0.55) & 0.003 \\
Dead animals are disposed of by refrigeration & 9 (1\%) & & 3 (\textless1\%) & & 0.59 & 0.7 (0.19-2.59) &  &  &  \\
Manure is stored using lagoon & 1,304 (82\%) & & 486 (55\%) & & \textless0.001 & 0.25 (0.21-0.30) &  &  &  \\
Manure is stored using deep pit & 241 (15\%) & & 303 (34\%) & & \textless0.001 & 2.89 (2.37-3.51) &  & 2.04 (1.41-2.95) & \textless0.001 \\
Manure is stored using tank & 59 (4\%) & & 106 (12\%) & & \textless0.001 & 4.23 (3.03-5.91) &  & 1.73 (1.07-2.81) & 0.03 \\
Manure is stored using land application & 56 (4\%) & & 99 (11\%) & & \textless0.001 & 4.21 (2.99-5.92) &  & 7.11 (4.91-10.29) & \textless0.001 \\
Rodent movement discouraged using bait & 1,446 (91\%) & & 654 (74\%) & & \textless0.001 & 0.19 (0.15-0.25) &  & 0.21 (0.15-0.31) & \textless0.001 \\
Rodent movement discouraged using grounds-keeping & 187 (12\%) & & 225 (25\%) & & \textless0.001 & 2.54 (2.05-3.14) &  & 6.68 (4.25-10.50) & \textless0.001 \\
\bottomrule 
\end{tabular}%
}
\begin{tablenotes}[flushleft]
\footnotesize
 \item[\dag] \small Odds ratio per \textbf{1,000-animal increase} in capacity.
 \item[\ddag] \small Odds ratio per \textbf{1,000\,m\textsuperscript{2} increase} in LOS area.
 \item[\S] \small Odds ratio per \textbf{100-meter increase} in perimeter.
\end{tablenotes}
\end{threeparttable}
\label{tab:results}
\end{table}

\section{Discussion} \noindent
RABapp\texttrademark\ streamlined the creation of standardized, electronic SPS biosecurity plans, alleviating producer burdens and enabling rapid review and approval by U.S. SAHOs.
Additionally, RABapp\texttrademark\ enhanced animal, semen \citep{cardenas_analyzing_2024}, and vehicle \citep{galvis_role_2024} traceability, tracked infection statuses, and integrated PRRSV, PEDV, and ASF transmission models \citep{sykes_estimating_2023, galvis_estimating_2025, galvis_betweenfarm_2022}.
The development and ongoing codification of the United States Swine Health Improvement Plan \href{https://www.aphis.usda.gov/livestock-poultry-disease/swine/us-ship}{(US SHIP)} \citep{harlow_biosecurity_2024} increased the adoption of SPS biosecurity plans among U.S. commercial swine producers due to US SHIP certification requirements \href{https://www.aphis.usda.gov/livestock-poultry-disease/swine/us-ship/enroll-certify}{(US SHIP-Enroll and Certify)}, which at the time of writing of this manuscript include: i) SPS biosecurity plan and ii) digital animal movement records.
By 2025, nearly 10,000 premises, predominantly finisher operations, had adopted RABapp\texttrademark, representing 31 U.S. states and 64 production companies. Analyzing RABapp\texttrademark repository, we highlighted regional disparities in biosecurity plan adoption (biosecurity deserts), described substantial variation in farm characteristics and practices, and identified SPS plan practices associated with PRRSV and PEDV infection risk: rendering, deep pit or tank manure storage, land application of manure, and grounds-keeping for rodent control were associated with higher infection odds, while footwear/clothing changes, greater distance to neighboring farms, carcass burial, rodent baiting, and multiple disposal sites were protective.
These findings underscored the potential of digital platforms like RABapp\texttrademark\ to strengthen disease prevention and response across the U.S. swine industry.

The total swine capacity of RABapp\texttrademark\ premises represented 42\% of the U.S. commercial hog inventory \citep{united_states_department_of_agriculture_2022_2024}. However, many swine-producing regions still have fewer than half of their hog population represented in RABapp\texttrademark\ (Figure \ref{fig:deserts}).
Of the ten ASDs with the largest swine inventories, seven were classified as biosecurity deserts. 
In biosecurity deserts, the limited adoption of RABapp\texttrademark\ suggests a lack of standardized biosecurity, which creates data gaps and could be considered a challenge to US SHIP national adoption, slowing the ability of state and federal officials to coordinate a rapid response to FADs.
Additionally, some SAHOs may require SPS plans to approve interstate movement permits, particularly for high-impact FADs such as ASF \citep{lee_swine_2021, usda-aphis_fad_2020}.
The absence of standardized plans may also increase the risk of infection with diseases such as PRRSV and PEDV, thereby contributing to the current circulation of endemic diseases.
For instance, our result showed higher outbreak frequencies in biosecurity deserts (p \ < 0.001), which indicated that such areas could benefit from developing or enhancing their current biosecurity.
We remark that the need for extension activities should prioritize biosecurity desert regions, where targeted outreach and incentive programs could address data gaps, potentially reducing endemic disease circulation and strengthening preparedness for FAD events.

We evaluated the written section of 5,515 SPS biosecurity plans and nearly all farms required footwear or clothing changes at entry (Figure \ref{fig:spsbar}), consistent with the 99.5\% reported by Harlow et al. \citep{harlow_biosecurity_2024}, suggesting strong industry consensus on the importance of reducing risk of fomite-driven disease introduction.
Carcass disposal methods showed greater heterogeneity. Rendering was included in 59\% of SPS plans across all production types, while composting (35\%), incineration (8\%), and burial (5\%) were less common.
The relatively high frequency of rendering contrasts with a 2012 Canadian study, which found that rendering was used on 43–49\% of sow and growing pig farms \citep{lambert_epidemiological_2012-1}. This difference may reflect an increased reliance on rendering services over time in the U.S., or regional differences in producer perceptions of disease risk and available infrastructure \citep{pudenz_adoption_2019, harlow_biosecurity_2024}.
Among sow farms, composting (50\%) was reported more frequently than rendering (41\%) (Supplementary Table S2), indicating that disposal preferences vary by production type.
Because breeding herds contain higher-value animals, outbreaks carry greater economic consequences, which may drive sow operations to adopt different carcass practices perceived as more biosecure.
The use of lagoons was by far the most used manure management method in southern regions (Figure \ref{fig:4panel}) and on sow (89\%), boar stud (83\%), and farrow-to-finish (80\%) premises, whereas deep pits were more prevalent in northern regions and on wean-to-finish (69\%), gilt (60\%), and isolation (90\%) premises (Supplementary Table S2). 
These patterns are consistent with climatic constraints, as lagoons function optimally in warm climates, while deep pits provide insulation against freezing temperatures in colder areas \citep{hatfield_swine_1998}.
Interestingly, breeding premises in the north more frequently reported lagoon storage than the growing-pig premises in the north, possibly reflecting operational differences in facility design and biosecurity priorities.
Carcass disposal methods also varied regionally, with rendering more common in the western U.S. and southeastern North Carolina, while composting predominated in eastern Iowa, Missouri, and much of the eastern U.S.
Economic and regulatory pressures also shape these patterns: rendering, while historically viewed as the most environmentally sound and biosecure option, has become more costly due to rising service fees and changing regulations \citep{kirstein_rendering_2002}. The relative cost differences between rendering and its alternatives likely vary geographically depending on local infrastructure and service availability.
Because we demonstrated that rendering was strongly associated with higher odds of PRRSV/PEDV infection in our regression analysis, its regional concentration underscores the need to consider geographic context when tailoring biosecurity recommendations.
The observed patterns suggest that while climate and infrastructure shape the feasibility of disposal and storage methods, producer choices within these constraints remain critical to disease risk.


Farm characteristics and biosecurity practices have a significant impact on PRRSV and PEDV risk nationwide.
A greater distance to the nearest premises was associated with reduced odds of PRRSV and PEDV occurrence. This relationship is consistent with the literature identifying proximity as a key risk factor for local transmission of PRRSV \citep{lambert_epidemiological_2012-1,sanchez_spatiotemporal_2023,mortensen_risk_2002,arruda_aerosol_2019,galvis_modelling_2022} and PEDV \citep{beam_porcine_2015,alvarez_spatial_2016,galvis_modeling_2022}.
Additionally, our study included two proximity-related variables: the distance to the nearest premises and the number of premises within a 10 km radius, which was chosen for its epidemiological relevance, as it captures potential lateral spread events by encompassing both short-distance farm-to-farm interactions \citep{kanankege_adapting_2022, galvis_betweenfarm_2022}.
Of these two proximity related variables, only the distance to the nearest premises was significant in the final multivariable model, suggesting that direct proximity to other farms may be more relevant than the local density of farms.
While the inclusion of either proximity-related variable in the final model allows it to capture the role of local transmission, the nearest-neighbor distance is preferred when analyzing risk or fitting disease transmission models to data.

We demostrated that deep-pit manure storage and land application of manure were both associated with higher odds of PRRSV/PEDV. 
In a study analyzing wean-to-market sites, recent manure pumping was linked to PRRSV outbreaks, with 3.38 times higher odds of an outbreak for pig lots that experienced pumping in the prior five weeks, and 3.38 times higher odds for lots when there was an outbreak in the previous lot, consistent with residual contamination and re-circulation during manure agitation and application \citep{serafini_poeta_silva_associations_2025}.
The same study reported lower odds of PRRSV in deep-pit barns than in lagoon or vat systems (OR = 0.35), likely reflecting confounding by site type because nurseries in that dataset used lagoon or vat storage and had higher outbreak risk; our multivariable results, which adjust for production type, instead show higher odds for deep pits and tanks \citep{serafini_poeta_silva_associations_2025}.
Because manure in deep-pit systems is stored beneath the barn, a persisting source of infection in the manure may be spread in and around the building by aerosols created by agitation and pumping.
During land application, agitation and spray generate aerosols, and shared hoses, tanks, and contractor equipment moving among sites act as fomites; these routes and the higher visit counts from multiple contractors align with the strong association between land application and increased risk seen in our results \citep{serafini_poeta_silva_associations_2025}.
While storing manure in tanks or concrete structures may be an alternative to deep pits in areas where climatic constraints make lagoons less feasible, our results showed that these alternatives also have an association with increased risk, suggesting that they may not confer an advantage.
The recent study on manure pumping and land application observed no statistical association between manure practices and PEDV, likely due to few PEDV events following pumping; therefore, our combined PRRSV/PEDV outcome may be primarily driven by PRRSV in the context of manure activities \citep{serafini_poeta_silva_associations_2025}.
Together, these findings support that manure storage and land application may be linked PRRSV transmission, and reinforce the need for biosecurity around manure pumping events, equipment sanitation, and contractor logistics.

Rendering carcasses significantly increased the risk of outbreaks to more than three times that of premises not using rendering, aligning with previous studies by Lambert et al., who found that rendering truck access was significantly associated with PRRSV infection \citep{lambert_epidemiological_2012}, and by Velasova et al., who reported higher PRRSV infection risks associated with dead pig collection \citep{velasova_risk_2012}. 

While SPS biosecurity plan maps are not required to have the carcass disposal (CD) located outside of the perimeter buffer area (PBA), encouraging this practice for premises that use rendering could potentially help to mitigate some of this risk \citep{alarcon_biosecurity_2021}.

Our results also showed that biosecurity plans stipulating footwear or clothing change requirements were linked to reduced outbreak risks, consistent with documented evidence of the rapid indirect transmission of PEDV via contaminated personal protective equipment \citep{kim_evaluation_2017} and the critical role of hygiene and barrier entry procedures in limiting disease spread \citep{harlow_biosecurity_2024}.

Rodent control measures were strongly associated with PRRSV/PEDV risk, consistent with evidence that rodents can act as vectors for swine pathogens, including PRRSV \citep{abdisa_serbessa_review_2023, lambert_epidemiological_2012-1}. The use of bait, typically in bait boxes, was the most common strategy (reported by 93\% of premises) and was strongly protective (OR: 0.21, CI: 0.15-0.31, p \textless\ 0.001). In contrast, farms reporting mowing or vegetation clearing around barns as a rodent control method had substantially higher odds of infection. A likely explanation is that mowing and grounds-keeping are often carried out by external contractors who visit multiple farms, creating opportunities for pathogen transfer via contaminated equipment, vehicles, or clothing \citep{abdisa_serbessa_review_2023, lambert_epidemiological_2012}. In addition, a personal communication with one swine veterinarian noted that contractors sometimes schedule visits in order of herd health status to reduce this risk; however, this approach depends on timely detection and cannot fully prevent indirect transmission. Importantly, grounds-keeping was never reported as the sole rodent control method; it was always used in conjunction with bait. This suggests that producers view bait as the primary strategy and vegetation management as a supplementary approach. Therefore, the increased risk is unlikely due to farms choosing grounds-keeping instead of bait, but rather to the added external contacts introduced by the grounds-keeping activity itself. This interpretation is consistent with previous studies showing that human and vehicle movements, whether for rendering, manure handling, feed delivery, or other activities, can inadvertently serve as pathways for disease spread \citep{lambert_epidemiological_2012-1,serafini_poeta_silva_associations_2025}. Thus, while rodent control is an important component of farm biosecurity, the method matters: approaches that directly reduce rodent populations (e.g., baiting) appear protective, whereas those that involve additional outside labor and equipment (e.g., mowing) may paradoxically increase disease risk.

Our findings demonstrate the potential of integrated digital tools like RABapp\texttrademark\ to not only support national programs but also to analyze large amounts of data to examine disease occurrence risks. As adoption grows, so will the potential for data-driven recommendations that are targeted, tailored, and timely. Producers, veterinarians, regulators, and other stakeholders all stand to gain from the enhanced transparency, traceability, and continuous improvements in biosecurity brought by RABapp\texttrademark. Ultimately, this study highlights the importance of collaborative ventures that combine epidemiological insight with practical, operational benefits.

\section{Limitations and final remarks}
\noindent This study has limitations related to the completeness of biosecurity and disease occurrence data in the RABapp\texttrademark\ repository. Most RABapp\texttrademark\ premises were managed by vertically integrated swine companies that were already invested in biosecurity, well-equipped to engage with national programs, and had recently been required to have SPS plans by US SHIP certification requirements. Independent and niche-market producers were underrepresented. While independent producers may not be aware of the necessity of SPS plans, they were able to enter plans in RABapp\texttrademark. Niche-market producers, with pigs with outdoor access, were not required to have plans for US SHIP certification, and RABapp\texttrademark\ did not provide support for such groups in entering their plans; thus, our results should not be interpreted as representing niche-market biosecurity.
Given the known risks of cross-system transmission, including contact with feral swine and outdoor pigs \citep{pepin_risk_2023}, extending RABapp\texttrademark\ to include outdoor, niche-market, and exhibition producers will enhance national biosecurity levels.

Our unit of analysis was the premises PIN level. Some premises shared the same coordinates (4.2\% of all premises; 10.1\% of the modeled subset), indicating multiple PINs at a single location.
Treating each PIN as an independent unit may have lowered between-farm distance metrics, potentially biasing estimates of spatial distances we used in our regression analysis.

Our estimates of biosecurity coverage used the 2022 Census of Agriculture inventory data, which reflects the population as of December 31, 2022 \citep{united_states_department_of_agriculture_2022_2024}. By contrast, RABapp\texttrademark\ premises report their capacity, which is the maximum number of animals the premises can hold.
Mismatch between the point-in-time inventory in the Census of Agriculture and reported capacity in RABapp\texttrademark\ could affect district-level coverage estimates.
Despite this, the 50\% coverage threshold remains a practical benchmark for identifying biosecurity deserts.

SPS biosecurity plans are intended for FAD events, not necessarily routine practices, and thus may not be implemented before needed. Additionally, we extracted variables from free-text response fields in the SPS biosecurity plans using string matching. Despite manual review, variations in terminology across regions, companies, and production types may have impacted the accuracy of the variables created.

The collection of PRRSV and PEDV outbreak data depends on heterogeneous testing and reporting.
Testing for either pathogen is driven mainly by outbreak investigations among growing pig farms, systematic testing among breeding farms, or the surveillance policies of a production company \citep{perez_individual_2019}. While under-reporting could have biased our results, prior studies suggest a strong correlation between reported cases in breeding herds and true field conditions \citep{galvis_modelling_2022,serafini_poeta_silva_monitoring_2024,sanchez_spatiotemporal_2023}.
Still, disease reporting from growing pig farms is a well-known area for future improvements, especially when disease elimination is of interest to producers.

Despite these limitations, the scope and depth of the RABapp\texttrademark\ repository offer unique opportunities. Although efforts are being employed across the U.S. to improve on-farm biosecurity and traceability for both endemic and anticipated foreign animal disease outbreaks, these efforts are limited due to their scope, often being done at a regional or state level.
With nearly half of the commercial swine population represented, RABapp\texttrademark\ is uniquely positioned for national-scale impact. The on-farm biosecurity, traceability, and mathematical modeling power of RABapp\texttrademark\ can be used to identify on-farm biosecurity practices and movement restrictions most effective in reducing local infection pressures and preventing the transmission of pathogens from growing-pig farms to proximal breeding farms, thereby significantly improving the U.S. swine industry’s ability to prevent and respond to endemic, emerging, and re-emerging high-impact swine diseases. Another valuable aspect of RABapp\texttrademark\ is its GIS-enabled functionality. Through interactive maps, outbreak simulations, and data-rich visualizations, RABapp\texttrademark\ helps producers and government officials make decisions about movements and biosecurity before and during emergency events.

\section{Conclusion} \noindent
This study evaluated swine farm biosecurity across 31 U.S. states and identified specific practices associated with reduced or increased risk of PRRSV and PEDV occurrence, using data from 7,781 premises and 5,515 SPS biosecurity plans on RABapp\texttrademark.
We found significant variation in SPS plans and biosecurity, especially across production types, and identified biosecurity deserts---regions where low RABapp\texttrademark\ adoption may hinder outbreak response and increase vulnerability. In Germany, farmers receive compensation if they meet legal biosecurity standards \citep{klein_biosecurity_2024}. A similar "compensation-for-compliance" policy in the U.S. could incentivize the broader adoption of SPS plans. This approach would complement existing programs, such as US SHIP, which already require biosecurity plans for certification of commercial swine farms.

We demonstrated that requirements for footwear or clothing changes, greater distance to neighboring farms, multiple carcass disposal locations, and use of bait to discourage rodent movement reduced the risk of PRRSV and PEDV infection. Conversely, using rendering for carcass disposal, deep-pit storage, tank storage, or land application of manure, and grounds-keeping to discourage rodent movement increased the risk of PRRSV and PEDV infection. These findings support broader SPS plan adoption and centralized data platforms like RABapp\texttrademark\ as essential tools. Wider adoption will enhance coordination among producers and regulators and strengthen the U.S. swine industry's ability to detect, respond to, and prevent disease outbreaks.

\subsubsection*{Acknowledgments}
\noindent We want to thank all participating swine producers and integrators for providing data and constructive feedback to RABapp\texttrademark. The role of the 31 departments of agriculture that use RABapp\texttrademark\ and provide insightful user feedback needs to be acknowledged. Finally, we thank the Morrison Swine Health Monitoring Program (MSHMP) for sharing disease outbreak data.


\subsubsection*{Funding}
\noindent This work was also supported by the Foundation for Food \& Agriculture Research (FFAR) award number FF-NIA21-0000000064. This projected was also funded by USDA's Animal and Plan Health Inspection Service via the National Animal Disease Preparedness and Response Program by a cooperative agreement under USDA-APHIS Award: AP22VSSP0000C004. The findings and conclusions in this document are those of the author(s) and should not be construed to represent any official USDA or U.S. Government determination or policy. This research received funding from the Swine Health Information Center (SHIC) under grant agreement number 23–028. This project is funded by National Institute of Food and Agriculture; Food and Agriculture Cyberinformatics and Tools, Grant/Award Numbers:2020-67021-32462 

\bibliographystyle{elsarticle-harv} 
\bibliography{references}
\end{document}


\begin{center}
    \large 
    \bfseries 
    Enhancing U.S. swine farm preparedness for infectious foreign animal diseases with rapid access to biosecurity information
    \par 
    \vspace{1.5em} 
\end{center}
\section{SPS biosecurity plan processing}

\noindent  SPS biosecurity plan information was retrieved from the RABapp\texttrademark\ repository and processed to prepare it for analysis. The processing described below applies only to the elements of the written component of the SPS biosecurity plan that were used in the analysis.

\paragraph{Protecting the pig herd}
The protecting the pig herd section includes this prompt: "The Perimeter Buffer Area (PBA) is labeled on the premises map at the end of this plan. The PBA is marked on-site with \_." Responses in this field containing "fence," "fencing," "rope," or "tape" were classified as implying a physical barrier to entry, as opposed to just a visual marker.

\paragraph{Carcass disposal}
The carcass disposal section includes this prompt: "In an FAD outbreak, dead animals (normal mortality numbers) are disposed of by \_ in a way which prevents the attraction of wildlife, rodents, and other scavengers, and is in accordance with state and federal laws." Responses in this field were classified into five non-exclusive categories by the presence of certain terms:
\begin{itemize}
    \item \textbf{Composting:} responses containing "compost," "comost" (misspelling of compost), "biovat," "dead box," "dead shed," or "adu box"
    \item \textbf{Incineration:} responses containing "incinerat"
    \item \textbf{Rendering:} responses containing "render"
    \item \textbf{Burial:} responses containing "burial" or "bury"
    \item \textbf{Refrigeration:} responses containing "refrigerated" or "cold box"
\end{itemize}

\noindent The next field is related to whether dead animals are disposed of within the PBA (Yes/No). The following field asks for the method used, with "N/A" (not applicable) for premises that responded with "No" in the previous field. Other responses were classified using the same method as the first prompt. The results were combined, such that the carcass disposal methods of a premises represented all categories found in either field.

\paragraph{Manure management}
The manure management section includes this prompt: "Manure is stored \_." Responses were classified into four non-exclusive categories by the presence of certain terms:
\begin{itemize}
    \item \textbf{Lagoon:} responses containing "lagoon," "laguon," "lagaon," "pond," or "earthen"
    \item \textbf{Pit:} responses containing "pit"
    \item \textbf{Tank:} responses containing "tank," "concrete," or "slurry"
    \item \textbf{Spray:} responses containing "spray," "field," or "broadcast"
\end{itemize}

\paragraph{Pest control}

\noindent The pest control section includes this prompt: "This site utilizes \_ to discourage rodent movement." Responses were classified into two non-exclusive categories by the presence of certain terms:
\begin{itemize}
    \item \textbf{Bait:} responses containing "bait" or "rotation"
    \item \textbf{Grounds-keeping:} responses containing "barrier," "perimeter," or "mow"
\end{itemize}

\noindent Another field in this section talks about whether bird netting is used to keep birds out of the building (Yes/No). The next field asks if buildings are totally enclosed to prevent bird entry (Yes/No). These responses were combined into a single variable indicating a response of "Yes" in either field.

\paragraph{Personnel}
The personnel section includes fields asking if individuals crossing the LOS must shower in and out, and if they must put on site-specific coveralls, clothing, and footwear at the LOS Access Point (Yes/No). For sites with multiple pig buildings without areas for changing into site-specific clothing, there are prompts for the steps taken, including a field asking if site-specific clothing or coveralls and footwear are put on and hands are washed when entering the PBA (Yes/No). These responses were combined into a single variable indicating a response of "Yes" in either field.

\paragraph{Animal and semen movement}
This section includes a field asking if animals are accepted from operations within an FMD Control Area (Yes/No).

\paragraph{Feed}
This section includes a field asking if feed trucks deliver feed without entering the PBA during an outbreak (Yes/No).

\section{Variables excluded during model selection}
\noindent  Of the initial 27 candidate predictors, nine were excluded for univariable $p$-values exceeding 0.20, and three were excluded due to multicollinearity (Table \ref{tab:removed_vars}). The final model was selected using backward elimination with $\alpha = 0.05$, removing predictors iteratively based on the highest $p$-value.

\begin{table}[h]
\centering
\caption{Candidate predictor variables excluded from the final multivariable model}
\begin{threeparttable}
\label{tab:removed_vars}
\resizebox{\textwidth}{!}{%
\begin{tabular}{lll}
\toprule
\textbf{Variable} & \textbf{Reason for exclusion} & \textbf{Details} \\
\midrule
Has cattle business & Univariable $p > 0.2$ & $p = 0.575$ \\
Has miscellaneous business & Univariable $p > 0.2$ & $p = 0.939$ \\
Dead animals are disposed of by refrigeration & Univariable $p > 0.2$ & $p = 0.593$ \\
Premises within 10 km & Univariable $p > 0.2$ & $p = 0.525$ \\
Capacity & Univariable $p > 0.2$ & $p = 0.218$ \\
Number of SE & Univariable $p > 0.2$ & $p = 0.337$ \\
Number of LOSAP & Univariable $p > 0.2$ & $p = 0.296$ \\
LOS area & Univariable $p > 0.2$ & $p = 0.914$ \\
LOS perimeter & Univariable $p > 0.2$ & $p = 0.467$ \\
\midrule
Buildings totally enclosed or netting used to keep birds out & Multicollinearity & Correlated with 'manure is stored using land application' ($\Phi$\textsuperscript{\dag} $= -0.97$) \\
Manure is stored using lagoon & Multicollinearity & Correlated with 'manure is stored using deep pit' ($\Phi$\textsuperscript{\dag} $= -0.80$) \\
Dead animals are disposed of by composting & Multicollinearity & Correlated with 'dead animals are disposed of by rendering' ($\Phi$\textsuperscript{\dag} $= -0.74$) \\
\midrule
Accepts animals from operations within an FMD control area & backward selection & $p = 0.913$ \\
Dead animals are disposed of by incineration & Backward selection & $p = 0.493$ \\
Number of LOS & Backward selection & $p = 0.220$ \\
Has crops business & Backward selection & $p = 0.180$ \\
Has poultry business & Backward selection & $p = 0.084$ \\
\bottomrule
\end{tabular}
}
\begin{tablenotes}[flushleft]
\footnotesize
 \item[\dag] \small Pearson $\chi^2$.
\end{tablenotes}
\end{threeparttable}
\end{table}

\section{SPS biosecurity plan information}
\noindent This section analyzes 5,515 Secure Pork Supply (SPS) biosecurity plans to characterize the on-farm biosecurity measures of premises hosted in RABapp\texttrademark\. We summarize the distribution of SPS plan features, while stratified by production type (Table \ref{tab:sps1}).

\begin{table}[!h]
\caption{Summary of selected SPS biosecurity plan information by production type for 5,515 RABapp\texttrademark\ premises}
\label{tab:sps1}
\resizebox{\textwidth}{!}{%
\begin{tabular}{llllllllll}
\hline
& All ($N=5,515$) &  & Sow ($N=613$) &  & Finisher ($N=2,302$) &  & Boar stud ($N=29$) &  & Gilt ($N=48$) \\ \cline{2-2} \cline{4-4} \cline{6-6} \cline{8-8} \cline{10-10}
SPS biosecurity plan characteristic & n (\%) &  & n (\%) &  & n (\%) &  & n (\%) &  & n (\%) \\ \hline
Separate animal loading \& people entry points &  4,358 (79\%) &  & 602 (98\%) &  & 1,542 (67\%) &  & 28 (97\%) &  & 42 (88\%) \\
Buildings totally enclosed or netting used to keep birds out &  5,349 (97\%) &  & 589 (96\%) &  & 2,219 (96\%) &  & 29 (100\%) &  & 48 (100\%) \\
Feed trucks do not enter PBA &  3,979 (72\%) &  & 455 (74\%) &  & 1,802 (78\%) &  & 18 (62\%) &  & 29 (60\%) \\
Accepts animals from operations within an FMD control area &  1,432 (26\%) &  & 172 (28\%) &  & 638 (28\%) &  & 6 (21\%) &  & 13 (27\%) \\
Perimeter buffer area (PBA) is marked on-site with a physical barrier &  4,193 (76\%) &  & 513 (84\%) &  & 1,807 (78\%) &  & 26 (90\%) &  & 38 (79\%) \\
Site-specific clothing or coveralls and footwear are put on &  5,306 (96\%) &  & 588 (96\%) &  & 2,167 (94\%) &  & 29 (100\%) &  & 48 (100\%) \\
Dead animals are disposed of by composting &  1,920 (35\%) &  & 307 (50\%) &  & 790 (34\%) &  & 6 (21\%) &  & 29 (60\%) \\
Dead animals are disposed of by incineration &  431 (8\%) &  & 148 (24\%) &  & 78 (3\%) &  & 19 (66\%) &  & 6 (12\%) \\
Dead animals are disposed of by rendering &  3,230 (59\%) &  & 254 (41\%) &  & 1,364 (59\%) &  & 3 (10\%) &  & 9 (19\%) \\
Dead animals are disposed of by burial &  262 (5\%) &  & 79 (13\%) &  & 105 (5\%) &  & 7 (24\%) &  & 5 (10\%) \\
Dead animals are disposed of by refrigeration &  13 (<1\%) &  & 6 (1\%) &  & 3 (<1\%) &  & 0 &  & 2 (4\%) \\
Manure is stored using lagoon &  2,651 (48\%) &  & 547 (89\%) &  & 1,154 (50\%) &  & 24 (83\%) &  & 20 (42\%) \\
Manure is stored using deep pit &  2,765 (50\%) &  & 104 (17\%) &  & 1,162 (50\%) &  & 5 (17\%) &  & 29 (60\%) \\
Manure is stored using tank &  188 (3\%) &  & 9 (1\%) &  & 3 (<1\%) &  & 0 &  & 0 \\
Manure is stored using land application &  159 (3\%) &  & 23 (4\%) &  & 82 (4\%) &  & 0 &  & 0 \\
Rodent movement discouraged using bait &  5,134 (93\%) &  & 613 (100\%) &  & 2,297 (100\%) &  & 29 (100\%) &  & 48 (100\%) \\
Rodent movement discouraged using grounds-keeping &  726 (13\%) &  & 64 (10\%) &  & 225 (10\%) &  & 3 (10\%) &  & 19 (40\%) \\
Vehicle movement (VM) inside PBA &  2,203 (40\%) &  & 265 (43\%) &  & 843 (37\%) &  & 13 (45\%) &  & 13 (27\%) \\
Designated parking area (DPA) inside PBA &  439 (8\%) &  & 94 (15\%) &  & 81 (4\%) &  & 4 (14\%) &  & 10 (21\%) \\
Carcass disposal (CD) inside PBA &  1,258 (23\%) &  & 161 (26\%) &  & 466 (20\%) &  & 12 (41\%) &  & 10 (21\%) \\ \hline
SPS biosecurity plan characteristic & Median (IQR) &  & Median (IQR) &  & Median (IQR) &  & Median (IQR) &  & Median (IQR) \\ \hline
Number of SE &  1 (1-1) &  & 1 (1-1) &  & 1 (1-1) &  & 1 (1-1) &  & 1 (1-1) \\
Number of LOSAP &  5 (3-8) &  & 8 (5-13) &  & 4 (3-7) &  & 5 (4-7) &  & 5 (3-6) \\
Number of PBAAP &  4 (2-6) &  & 5 (3-7) &  & 4 (2-6) &  & 3 (3-6) &  & 3 (2-4.25) \\
Number of LOS &  1 (1-2) &  & 1 (1-1) &  & 2 (1-3) &  & 1 (1-2) &  & 1 (1-1.25) \\
Number of CD &  1 (1-1) &  & 1 (1-1) &  & 1 (1-1) &  & 1 (1-1) &  & 1 (1-1) \\
Number of PBA &  1 (1-1) &  & 1 (1-1) &  & 1 (1-1) &  & 1 (1-1) &  & 1 (1-1) \\
Number of DPA &  1 (1-1) &  & 1 (1-1) &  & 1 (1-1) &  & 1 (1-1) &  & 1 (1-1) \\
Number of cleaning stations &  1 (1-1) &  & 1 (1-2) &  & 1 (1-2) &  & 1 (1-2) &  & 1 (1-1) \\
LOS area (m\textsuperscript{2}) &  3,143 (1,938-4,550) &  & 8,435 (6,035-12,991) &  & 3,110 (1,937-4,268) &  & 1,342 (1,101-2,148) &  & 2,077 (1,841-3,452) \\
LOS perimeter (m) &  467 (266-727) &  & 1,174 (821-1,856) &  & 470 (282-719) &  & 295 (179-377) &  & 380 (228-517) \\
PBA perimeter (m) &  428 (308-612) &  & 696 (541-870) &  & 406 (309-581) &  & 358 (277-472) &  & 402 (253-559) \\ \hline
\end{tabular}%
}
\end{table}

\begin{table}[H]
\ContinuedFloat
\caption{Summary of selected SPS biosecurity plan information by production type (Continued)}
\label{tab:sps2}
\resizebox{\textwidth}{!}{%
\begin{tabular}{llllllllll}
\hline
& Wean-to-finish ($N=1,751$) &  & Nursery ($N=685$) &  & Isolation ($N=18$) &  & Farrow-to-finish ($N=44$) &  & Other ($N=25$) \\ \cline{2-2} \cline{4-4} \cline{6-6} \cline{8-8} \cline{10-10}
SPS biosecurity plan characteristic & n (\%) &  & n (\%) &  & n (\%) &  & n (\%) &  & n (\%) \\ \hline
Separate animal loading \& people entry points & 1,444 (82\%) &  & 621 (91\%) &  & 18 (100\%) &  & 40 (91\%) &  & 21 (84\%) \\
Buildings totally enclosed or netting used to keep birds out & 1,731 (99\%) &  & 646 (94\%) &  & 18 (100\%) &  & 44 (100\%) &  & 25 (100\%) \\
Feed trucks do not enter PBA & 1,057 (60\%) &  & 567 (83\%) &  & 16 (89\%) &  & 15 (34\%) &  & 20 (80\%) \\
Accepts animals from operations within an FMD control area & 352 (20\%) &  & 211 (31\%) &  & 10 (56\%) &  & 18 (41\%) &  & 12 (48\%) \\
Perimeter buffer area (PBA) is marked on-site with a physical barrier & 1,172 (67\%) &  & 554 (81\%) &  & 18 (100\%) &  & 41 (93\%) &  & 24 (96\%) \\
Site-specific clothing or coveralls and footwear are put on & 1,747 (100\%) &  & 642 (94\%) &  & 18 (100\%) &  & 42 (95\%) &  & 25 (100\%) \\
Dead animals are disposed of by composting & 502 (29\%) &  & 225 (33\%) &  & 15 (83\%) &  & 29 (66\%) &  & 17 (68\%) \\
Dead animals are disposed of by incineration & 40 (2\%) &  & 116 (17\%) &  & 4 (22\%) &  & 9 (20\%) &  & 11 (44\%) \\
Dead animals are disposed of by rendering & 1,188 (68\%) &  & 389 (57\%) &  & 1 (6\%) &  & 4 (9\%) &  & 18 (72\%) \\
Dead animals are disposed of by burial & 37 (2\%) &  & 15 (2\%) &  & 1 (6\%) &  & 2 (5\%) &  & 11 (44\%) \\
Dead animals are disposed of by refrigeration & 1 (<1\%) &  & 1 (<1\%) &  & 0 &  & 0 &  & 0 \\
Manure is stored using lagoon & 373 (21\%) &  & 471 (69\%) &  & 6 (33\%) &  & 35 (80\%) &  & 21 (84\%) \\
Manure is stored using deep pit & 1,215 (69\%) &  & 216 (32\%) &  & 17 (94\%) &  & 16 (36\%) &  & 1 (4\%) \\
Manure is stored using tank & 167 (10\%) &  & 6 (1\%) &  & 0 &  & 2 (5\%) &  & 1 (4\%) \\
Manure is stored using land application & 15 (1\%) &  & 39 (6\%) &  & 0 &  & 0 &  & 0 \\
Rodent movement discouraged using bait & 1,379 (79\%) &  & 683 (100\%) &  & 18 (100\%) &  & 43 (98\%) &  & 24 (96\%) \\
Rodent movement discouraged using grounds-keeping & 345 (20\%) &  & 46 (7\%) &  & 11 (61\%) &  & 13 (30\%) &  & 0 \\
Vehicle movement (VM) inside PBA & 773 (44\%) &  & 266 (39\%) &  & 2 (11\%) &  & 25 (57\%) &  & 3 (12\%) \\
Designated parking area (DPA) inside PBA & 217 (12\%) &  & 27 (4\%) &  & 0 &  & 6 (14\%) &  & 0 \\
Carcass disposal (CD) inside PBA & 465 (27\%) &  & 125 (18\%) &  & 0 &  & 16 (36\%) &  & 3 (12\%) \\ \hline
SPS biosecurity plan characteristic & Median (IQR) &  & Median (IQR) &  & Median (IQR) &  & Median (IQR) &  & Median (IQR) \\ \hline
Number of SE & 1 (1-1) &  & 1 (1-1) &  & 1 (1-1) &  & 1 (1-1) &  & 1 (1-1) \\
Number of LOSAP & 5 (3-7) &  & 4 (3-5) &  & 2 (2-3.75) &  & 11 (6.5-15) &  & 7 (5-9) \\
Number of PBAAP & 4 (2-7) &  & 3 (2-5) &  & 3 (2-3.75) &  & 3 (2-6) &  & 4 (3-6) \\
Number of LOS & 1 (1-2) &  & 1 (1-1) &  & 1 (1-1) &  & 1 (1-2) &  & 1 (1-1) \\
Number of CD & 1 (1-1) &  & 1 (1-1) &  & 1 (1-2) &  & 1 (1-3) &  & 1 (1-1) \\
Number of PBA & 1 (1-1) &  & 1 (1-1) &  & 1 (1-1) &  & 1 (1-1.25) &  & 1 (1-1) \\
Number of DPA & 1 (1-1) &  & 1 (1-1) &  & 1 (1-1) &  & 1 (1-2) &  & 1 (1-1) \\
Number of Cleaning stations & 1 (1-1) &  & 1 (1-1) &  & 1 (1-1) &  & 1 (1-2) &  & 1 (1-1) \\
LOS area (m\textsuperscript{2}) & 3,218 (2,024-4,014) &  & 1,553 (956-2,297) &  & 843 (362-894) &  & 10,197 (8,478-12,876) &  & 2,622 (2,315-2,780) \\
LOS perimeter (m) & 444 (242-590) &  & 264 (145-481) &  & 147 (95-217) &  & 1,377 (971-1,978) &  & 391 (286-589) \\
PBA perimeter (m) & 453 (321-582) &  & 346 (195-402) &  & 169 (122-234) &  & 981 (823-1,180) &  & 342 (309-447) \\ \hline
\end{tabular}%
}
\end{table}